\documentclass[article]{JFM-FLM_Au}

\usepackage{amssymb}
\usepackage{graphicx}
\usepackage{bm}
\usepackage{physics}

\lefttitle{Y.\ Hosaka and K.\ Ishimoto}
\righttitle{Journal of Fluid Mechanics}

\title{Swimming formula in a linearly viscous fluid with anisotropic and odd viscosity tensors at low Reynolds number}

\author{Yuto Hosaka\aff{1}, Kenta Ishimoto\aff{1}}

\affiliation{\aff{1}Department of Mathematics, Kyoto University, Kyoto 606-8502, Japan}

\corresau{Yuto Hosaka, \email{hosaka.yuto.7r@kyoto-u.ac.jp}}

\begin{document}
\maketitle

\begin{abstract}
We theoretically investigate the low-Reynolds-number hydrodynamics of a microswimmer in a linearly viscous fluid.
The fluid is characterized by a viscosity tensor containing both even and odd components, and can therefore exhibit anisotropic and chiral properties absent in an isotropic Newtonian fluid.
By employing the generalized Lorentz reciprocal theorem, we establish the Onsager-Casimir reciprocity for the resistance tensors of an arbitrarily shaped rigid body. 
We then derive a swimming formula that linearly relates the swimming velocities to the prescribed surface velocity of a microswimmer and imposed background flows.
As a showcase application of the derived swimming formula, we prove that the scallop theorem holds in this class of linearly viscous fluids.
Furthermore, we investigate the equations of motion for a spherical swimmer in a fluid with odd viscosity, as a representative example of the generalized linear constitutive framework.
In particular, we reveal that sufficiently strong shear flows destabilize the bimodal chirotactic alignment of a microswimmer.
Our results highlight how intrinsic fluid chirality and anisotropy fundamentally alter the orientational dynamics and transport phenomena of microswimmers.
\end{abstract}

\section{Introduction}
\label{sec:intro}

Microorganisms frequently navigate complex fluids ranging from biological to synthetic structured media~\citep{spagnolie2023swimming}.
Unlike a simple isotropic Newtonian fluid, many such media possess a preferred microstructural orientation, giving rise to anisotropic mechanical and rheological responses.
In biological contexts, examples of structured biological environments include the extracellular matrix, biopolymer networks, and cervical mucus, whose fibrous microstructure motivates anisotropic models of microorganism transport \citep{cupples2017viscous, shi2017swimming, Cupples2018}. 
Similar axial ordering also arises in synthetic systems, such as liquid crystals \citep{aranson2018harnessing}, aligned colloidal suspensions \citep{snezhko2016complex}, actuated magnetic particles \citep{junot2023large, beppu2024magnetically}, and chiral columnar crystals \citep{Kole, maitra2025activity}.
Understanding how orientationally ordered media affect microswimmer motion and particle transport is important for reproductive biology, targeted drug delivery, and the design of microfluidic devices \citep{ju2025technology}.

To describe the mechanics of such structured media, continuum theories based on rational mechanics have been developed \citep{truesdell1992first, truesdell2004non}.
Extending the concept of simple materials established in the framework of rational mechanics, Ericksen formulated the fundamental continuum theory for fluids with uniaxial orientational order, known as transversely isotropic fluids \citep{ericksen1960transversely}. 
Under strict invariance conditions required of classical continua, this constitutive relation is uniaxial, characterizing the fluid anisotropy through a single preferred direction that represents the local microstructural orientation.
This framework was later extended into the hydrodynamic theory of nematic liquid crystals \citep{leslie1966some}.
Such continuum descriptions have served as a cornerstone for studying complex media, including fiber-reinforced fluids, polymeric suspensions, and the motion of microswimmers in liquid crystals \citep{daddi2018, lin2022anisotropic, mandal2025cooperativity}.
Numerical approaches, including multiparticle collision dynamics, have subsequently been developed to resolve the coupled flow and orientational dynamics \citep{mandal2019multiparticle, mandal2021multiparticle}.
These descriptions provide a foundation for anisotropic transport, while internally driven chiral media introduce additional constitutive couplings.

In recent years, research on active matter has extended this picture to systems composed of individual units that consume energy locally to perform mechanical work \citep{marchetti2013hydrodynamics, gompper20252025}.
A particularly intriguing subclass comprises chiral active fluids, where active agents break both time-reversal and parity symmetries \citep{furthauer2012active, liebchen2022chiral, hosaka2022nonequilibrium, fruchart2023odd, mecke2024emergent}. 
Examples of chiral active fluids range from chiral bacterial vortices \citep{beppu2021, nishiguchi2025vortex}, living chiral crystals \citep{tan2022odd}, and cilia-driven tissue surfaces to synthetic micro-rotors and actuated colloidal particles \citep{soni2019odd, lopez2022chirality, mecke2023simultaneous, chen2024self}. 
In suspensions of driven rotors, internal angular momentum and active torques can generate additional contributions to the macroscopic stress.
This coupling between directional order and microscopic chirality fundamentally alters the macroscopic hydrodynamic equations, giving rise to unconventional mechanical behaviors absent in classical fluids \citep{shankar2022topological}.

The combination of internal driving, chirality, and axial anisotropy has revealed a spectrum of novel transport phenomena.
A prominent consequence is the emergence of odd viscosity, a nondissipative contribution to the viscous stress \citep{avron1998, banerjee2017}.
Odd viscosity is the part of the viscosity tensor that is antisymmetric under exchange of the stress and strain-rate index pairs.
Its associated stress makes no direct contribution to the local viscous dissipation, and the odd viscosity therefore gives rise to transverse hydrodynamic responses, unlike the conventional shear viscosity of a Newtonian fluid.
This transverse property leads to a range of dynamical phenomena, and recent experimental and theoretical findings have demonstrated that chiral active fluids exhibit topological edge modes \citep{souslov2019topological}, unusual interfacial dynamics \citep{soni2019odd, abanov2020hydrodynamics, francca2026odd}, and nonreciprocal hydrodynamic responses, including asymmetric Green's functions for point-force responses \citep{hosaka2021nonreciprocal} and lift forces \citep{hosaka2021hydrodynamic, lier2023lift, lier2024slip}.

\begin{figure}[htbp]
    \centering
    \includegraphics[width=0.5\textwidth]{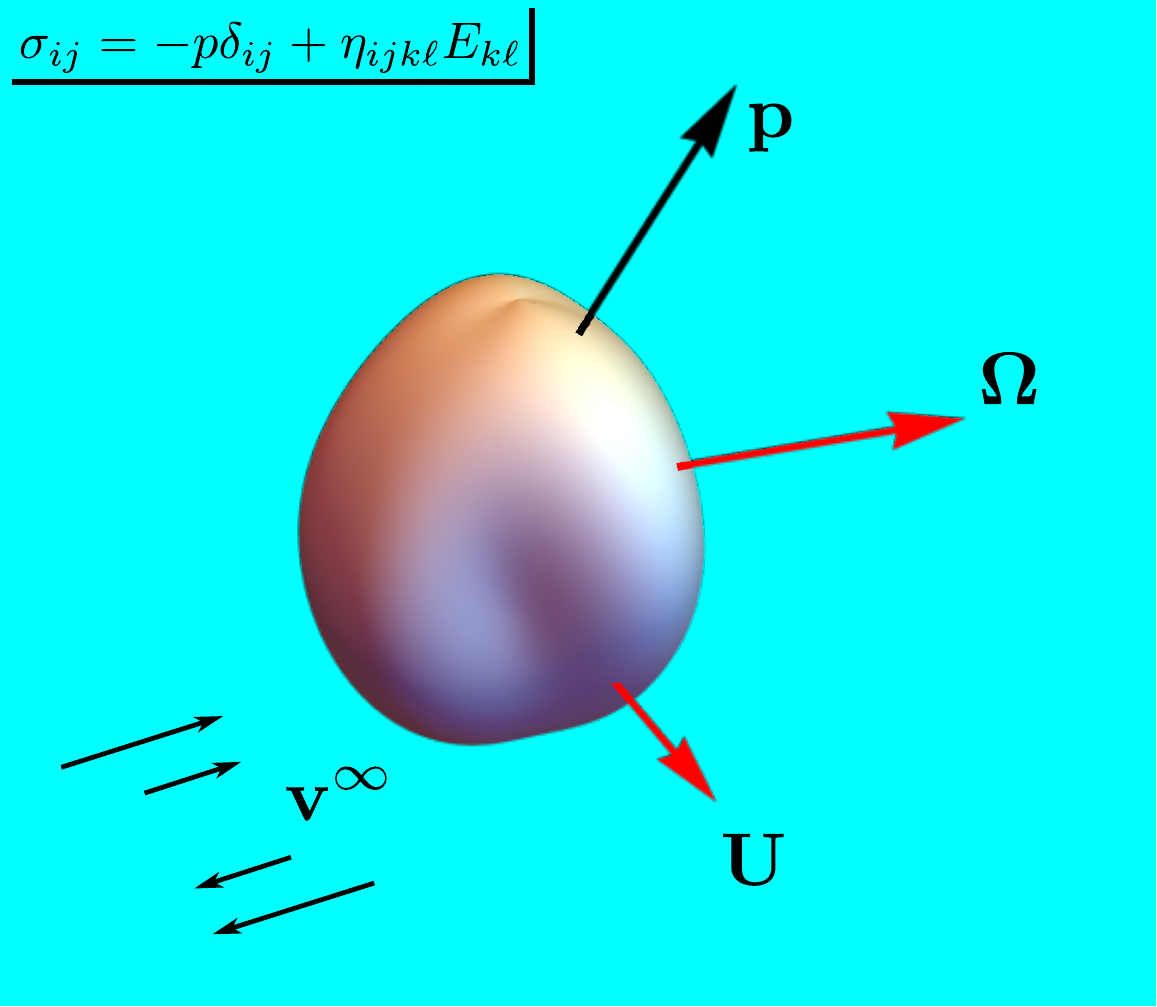}
    \caption{
Schematic illustration of an arbitrarily shaped microswimmer immersed in an unbounded, linearly viscous fluid subject to an imposed background shear flow $\mathbf{v}^{\infty}$.
The swimmer translates with velocity $\mathbf{U}$ and rotates with angular velocity $\boldsymbol{\Omega}$, while the body-fixed unit vector $\mathbf{p}$ specifies its orientation.
The fluid is characterized by the general constitutive relation $\sigma_{ij}=-p\delta_{ij}+\eta_{ijk\ell}E_{k\ell}$, where $p$ is the pressure, $E_{k\ell}$ is the rate-of-strain tensor, and $\eta_{ijk\ell}$ is a fourth-rank viscosity tensor without an assumed material symmetry [see equation~\eqref{eq:constitutive}].
    }
    \label{fig:setup}
\end{figure}

Although much of the early work on odd viscosity focused on two-dimensional systems, three-dimensional odd-viscous hydrodynamics has also advanced rapidly \citep{markovich2021}.
In particular, the hydrodynamics of spherical bodies in three-dimensional odd-viscous fluids has been investigated extensively \citep{khain2022, yuan2023stokesian,hosaka2023lorentz}, and exact solutions for a rigid sphere in models with a single odd viscosity coefficient have recently been obtained \citep{everts2024dissipative, MeissnerOszer.Everts2026}.
Related studies have also examined the motion of self-propelled particles and microswimmers in odd flows \citep{lapa2014, hosaka2023pair, hosaka2023hydrodynamics, hosaka2023lorentz, hosaka2024chirotactic, Chatzittofi2025}, as well as the sedimentation of particles with geometries beyond simple circular or spherical shapes \citep{khain2024trading}.
Despite these advances, particle transport in fluids described by more general viscosity tensors, including those with two independent odd viscosity coefficients, remains incompletely understood. 
A recent step in this direction was taken by Lier, who investigated particle transport in weakly odd-viscous fluids with two independent odd viscosity coefficients \citep{lier2026sorting}.
Beyond extending these studies to more general constitutive models, an important open question is how the surface deformations of an active particle or microswimmer determine its translational and rotational swimming velocities.
Surface integral formulas relating swimming velocities to prescribed surface deformations have been derived for Newtonian fluids \citep{stone1996propulsion, yariv2006self, ishimoto2023jeffery}, whereas their counterparts for fluids with anisotropic and odd viscosity tensors remain to be established.
Deriving such formulas requires an understanding of the symmetry relations of the hydrodynamic resistance tensors, whose explicit evaluation can be challenging for bodies of arbitrary shape. Related integral formulas for spherical swimmers have been discussed by one of the present authors \citep{hosaka2024chirotactic, chatzittofi2026self}. These developments lead to a natural question: how do anisotropic and odd contributions to the viscosity tensor modify the symmetry relations of particle resistance tensors and the resulting swimming formulas (figure \ref{fig:setup})?

In this study, we investigate the dynamics of an arbitrarily shaped body, which can be either a passive particle or an active swimmer, in a linearly viscous fluid characterized by a viscosity tensor.
Since the fluid under consideration is characterized by a generic viscosity tensor having anisotropic contributions, the presented theory can be applied to a wide range of complex fluids encompassing passive anisotropic fluids and chiral active fluids with odd viscosities.
Employing the generalized Lorentz reciprocal theorem \citep{hosaka2023lorentz}, we establish the Onsager-Casimir reciprocity of the resistance tensors of a rigid body of arbitrary shape, which represents symmetry between these tensors at opposite values of their antisymmetric components.
We then derive a generalized swimming formula that couples the active surface velocity of a microswimmer and external background flows with a viscosity tensor containing both even and odd contributions.
As a showcase application, we show that the scallop theorem holds in such generalized linear fluids. 
Furthermore, we investigate the equations of motion for a spherical swimmer in chiral fluids, identifying symmetry classes that permit translation-rotation coupling for a spherical body.
We also analyze the effects of background flow and reveal that under sufficiently high shear rates, microswimmers no longer exhibit the bistable orientational behavior, known as bimodal \textit{chirotaxis} \citep{hosaka2024chirotactic}.
Our results highlight how intrinsic fluid chirality and structural anisotropy fundamentally alter the orientation and transport phenomena of microparticles.

The paper is organized as follows: We first introduce the generalized linear constitutive relation for chiral anisotropic fluids in section~\ref{sec:constitutive}. 
We then discuss in section~\ref{sec:OCR} the symmetry of the resistance tensor of objects of arbitrary shape in such fluids, using the generalized Lorentz reciprocal theorem.
Section~\ref{sec:swimming} covers the derivation of a swimming formula for an active object in a background shear flow.
We then in section~\ref{sec:scallop} discuss the scallop theorem under the assumptions of the present constitutive model.
In section~\ref{sec:resistance}, we construct possible components in the resistance tensors compatible with a fluid axis of the continuous rotational symmetry.
We then derive the equations of motion for a passive spherical body in chiral fluids in section~\ref{sec:dynamics}, and discuss representative classifications of anisotropic fluids based on the symmetry argument.
Section \ref{sec:passive} examines swimmer dynamics in a simple shear flow in the special case of a single odd viscosity.
Finally, we discuss the results and present our conclusions in section~\ref{sec:discussion}.

\section{Generalized linearly viscous fluid model}
\label{sec:constitutive}

We consider an incompressible fluid in three-dimensional space in the low-Reynolds-number regime. 
Within a linear relation between the viscous stress and the velocity gradient, the stress tensor is generally expressed with a fourth-rank viscosity tensor $\bm\eta$
\begin{align}
    \sigma_{ij}=-p\delta_{ij}+
    \eta_{ijk\ell} E_{k\ell}
    .
    \label{eq:constitutive}
\end{align}
Here, $p$ is the pressure, $\delta_{ij}$ is the identity tensor, and $\mathbf{E}=\frac{1}{2}[\nabla\mathbf{v}+(\nabla\mathbf{v})^\top]$ is the rate of strain tensor with $\top$ being the transposition of a matrix, where $\mathbf{v}$ is the fluid velocity that satisfies the incompressibility condition $\nabla\cdot\mathbf{v}=0$.
In the above, we have assumed that the stress tensor is compatible with
conservation of angular momentum and the objectivity for the velocity gradient tensor.
Here and throughout the paper, we use the Einstein summation convention for the repeated spatial indices, $i,j,k,\ell=\{1,2,3\}$.

At low Reynolds numbers and in the steady state, the hydrodynamic equations can be obtained from the force-balance equation
\begin{align}
    \partial_j\sigma_{ij}=0.
    \label{eq:forcebalance}
\end{align}
While the angular-momentum conservation and the objectivity require the fourth-rank viscosity tensor, $\eta_{ijk\ell}$, to be symmetric under the index permutations $i\leftrightarrow j$ and $k\leftrightarrow \ell$ respectively, the tensor may not be symmetric upon exchanging the pair of indices $ij$ and $k\ell$, namely, $\eta_{ijk\ell}\neq\eta_{k\ell ij}$.
One can therefore decompose the tensor into the symmetric and antisymmetric components, $\bm{\eta}=\bm{\eta}^{\rm e}+\bm{\eta}^{\rm o}$, where each component satisfies
\begin{align}
	\eta_{ijk\ell}^{\rm e} = \eta_{k\ell ij}^{\rm e},\quad
	\eta_{ijk\ell}^{\rm o} = -\eta_{k\ell ij}^{\rm o}.
	\label{eq:eta}
\end{align}
The viscosity tensor invariant under $ij\leftrightarrow k\ell$ is an even viscosity tensor contributing to the fluid energy dissipation $\sigma_{ij}\partial_jv_i$, while the tensor that changes its sign upon the index permutation represents an odd viscosity tensor.
In an isotropic fluid, the odd viscosity vanishes $(\bm{\eta}^{\rm o}=\mathbf{0})$, as odd viscosities in three-dimensional fluids are not compatible with isotropy \citep{khain2022}.
In the presence of preferred directions, fluids become anisotropic, and odd viscosities can arise when a fluid is characterized by at least one axis of chirality \citep{fruchart2023odd}.

\section{Onsager-Casimir reciprocity}
\label{sec:OCR}

We first show that the resistance tensors of an arbitrary-shaped rigid object, which should be symmetric in a Newtonian fluid, can have antisymmetric components in a fluid characterized by a viscosity tensor.
To this aim, we use the Lorentz reciprocal theorem extended to such generic media~\citep{hosaka2023lorentz} and apply it to determine the symmetries of resistance tensors for a particle in Stokes flow \citep{hinch1972note, masoud2019}.

Consider a single rigid particle of arbitrary shape moving in a viscous fluid characterized by the force-balance equation~\eqref{eq:forcebalance}.
As schematically illustrated in figure~\ref{fig:setup}, the particle moves with a linear velocity $\mathbf{U}$, an angular velocity $\bm{\Omega}$, and a linear background flow
\begin{align}
    \mathbf{v}^\infty(\mathbf{r}) 
    = \mathbf{U}^\infty + \boldsymbol{\Omega}^\infty\times\mathbf{r}
    +
    \mathbf{E}^\infty\cdot\mathbf{r}
    +
    \mathcal{O}(|\mathbf{r}|^2),
    \label{eq:infinity}
\end{align}
where $\mathbf{U}^\infty$ is the linear background velocity, $\boldsymbol{\Omega}^\infty$ is the background angular velocity, and $\mathbf{E}^\infty=\frac{1}{2}[\nabla\mathbf{v}^\infty+(\nabla\mathbf{v}^\infty)^\top]$ is the rate-of-strain tensor from the background flow with $\mathbf{r}$ being a position vector measured from the origin.
We have assumed here that the background flow field in the absence of the object is locally approximated by a linear flow, which satisfies the Stokes equations~\eqref{eq:forcebalance}.

The hydrodynamic force, torque, and stresslet tensor exerted by the fluid on the particle are, respectively, defined as $\mathbf{F}=-\int_\mathcal{S}dS\, \bm{\sigma}\cdot\mathbf{n}$, $\mathbf{M}=-\int_\mathcal{S}dS\, \mathbf{r}\times(\bm{\sigma}\cdot\mathbf{n})$, and $\mathbf{S}=-\frac{1}{2}\int_\mathcal{S}dS\,[\mathbf{r}(\bm{\sigma}\cdot\mathbf{n})+(\bm{\sigma}\cdot\mathbf{n})\mathbf{r}]$ for a rigid body with a no-slip boundary condition.
Here $\mathbf{n}$ denotes the outward unit normal from the fluid region into the particle, and the surface integrals are performed over the particle surface $\mathcal{S}$.
From the linearity of the governing equations, the force moments and the velocity tensor have to be linearly linked via the resistance tensors with the following nine tensorial subblocks
\begin{align}
    \begin{pmatrix}
    	\mathbf{F}\\    \mathbf{M} \\ \mathbf{S}
     \end{pmatrix}
    =
    -
    \begin{pmatrix}
        \mathbf{K}^{FU} & \mathbf{K}^{F\Omega} & \mathbf{K}^{FE} \\
        \mathbf{K}^{MU} & \mathbf{K}^{M\Omega} & \mathbf{K}^{ME} \\
        \mathbf{K}^{SU} & \mathbf{K}^{S\Omega} & \mathbf{K}^{SE} 
    \end{pmatrix}
    \begin{pmatrix}
    	\mathbf{U}-\mathbf{U}^\infty \\    \boldsymbol{\Omega}-\boldsymbol{\Omega}^\infty \\  {-\mathbf{E}^\infty}
     \end{pmatrix}
    .
    \label{eq:grand}
\end{align}
From the symmetric property of the stresslet tensor, without loss of generality, we can define $\mathbf{K}^{SU}, \mathbf{K}^{S\Omega},$ and $\mathbf{K}^{SE}$ so that they are symmetric with respect to their first and second indices, i.e., $K^{SU}_{ijk}=K^{SU}_{jik}, K^{S\Omega}_{ijk}=K^{S\Omega}_{jik}$, and $K^{SE}_{ijk\ell}=K^{SE}_{jik\ell}$.
Since $\mathbf{E}^\infty$ is symmetric, $\mathbf{K}^{FE}, \mathbf{K}^{ME},$ and $\mathbf{K}^{SE}$ may be chosen to be symmetric in their last two indices, i.e., $K^{FE}_{ijk}=K^{FE}_{ikj}, K^{ME}_{ijk}=K^{ME}_{ikj}$, and $K^{SE}_{ijk\ell}=K^{SE}_{ij\ell k}$.

For a Newtonian fluid with a shear viscosity alone, the resistance tensors of a rigid body with a no-slip boundary condition obey the symmetric relations, and the number of independent tensorial blocks reduces to six \citep{hinch1972note, masoud2019}.
We use the Lorentz reciprocal theorem extended to fluids characterized by the tensorial viscosity \citep{hosaka2023lorentz}.
Consider two Stokes flows without body-force densities: the main flow described by $\partial_j\sigma_{ij}=0$ and $\partial_iv_i=0$ and the auxiliary flow (denoted by the hat symbol $\hat{}\,$) described by $\partial_j\hat{\sigma}_{ij}=0$ and $\partial_i\hat{v}_i=0$, where $\sigma_{ij}=-p\delta_{ij}+\eta_{ijk\ell}\partial_\ell v_k$ and $\hat{\sigma}_{ij}=-\hat{p}\delta_{ij}+\hat{\eta}_{ijk\ell}\partial_\ell\hat{v}_k$.
Although in the previous section, we have assumed the symmetries of $\bm{\eta}$ and $\bm{\hat{\eta}}$ under $i\leftrightarrow j$ and $k\leftrightarrow \ell$, the results in this section apply to any viscosity tensor, and thus the stress and the velocity gradient tensors can be asymmetric.
For fluids with the generic viscosity tensors as in equation~\eqref{eq:eta}, the generalized Lorentz reciprocal theorem takes the form \citep{hosaka2023lorentz}:
\begin{align}
    \int_\mathcal{S} dS\,
    \mathbf{v}\cdot\hat{\boldsymbol{\sigma}}\cdot\mathbf{n}
    =
    \int_\mathcal{S} dS\,
    \hat{\mathbf{v}}\cdot\boldsymbol{\sigma}\cdot\mathbf{n},
    \label{eq:LRT}
\end{align}
under the conditions between the viscosity tensors in the main and auxiliary problems:
\begin{align}
    \bm{\eta}^{\rm e} = \bm{\hat{\eta}}^{\rm e},\quad
    \bm{\eta}^{\rm o} = -\bm{\hat{\eta}}^{\rm o}.
    \label{eq:cond}
\end{align}
In the above, the surface integrals are performed over the object surface $\mathcal{S}$.

To demonstrate the identities of the resistance tensors, we use the generalized reciprocal theorem \eqref{eq:LRT} and the condition \eqref{eq:cond} for the two corresponding motions.
As a showcase demonstration, we first consider two distinct translational motions of the particles, $\mathbf{U}$ and $\mathbf{\hat{U}}$ to derive the relation for $\mathbf{K}^{FU}$.
We then consider two motions involving a shear flow $\mathbf{E}^\infty$ and a rotation $\bm{\Omega}$, to derive the relation between $\mathbf{K}^{S\Omega}$ and $\mathbf{K}^{ME}$.
The identities involving the other resistance tensors can be proven similarly using the generalized reciprocal theorem.

For the reciprocal theorem applied to the first case with the force relations $\mathbf{F}=-\mathbf{K}^{FU}\cdot\mathbf{U}$ and $\mathbf{\hat{F}}=-\mathbf{\hat{K}}^{FU}\cdot\mathbf{\hat{U}}$, contributions at large distances vanish in an unbounded flow, because the velocity decays with the inverse of the distance, akin to the classical singularity \citep{everts2024dissipative, MeissnerOszer.Everts2026}.
The reciprocal theorem then reduces to the surface integral of the no-slip boundary, and takes the bilinear form of the velocities:
\begin{align}
\mathbf{U}\cdot\mathbf{\hat{K}}^{FU}\cdot\mathbf{\hat{U}}
=
\mathbf{\hat{U}}\cdot\mathbf{K}^{FU}\cdot\mathbf{U}.
\end{align}
Through the transformation $U_i(\hat{K}^{FU}_{ij}-K^{FU}_{ji})\hat{U}_j=0$, and given that the velocity vectors can be arbitrary, we find the relation $\hat{K}^{FU}_{ij}=K^{FU}_{ji}$.
Applying Conditions~\eqref{eq:cond} to this relation between the two problems, we conclude the symmetry of the force-velocity tensor
\begin{align}
K_{ij}^{FU}(\bm{\eta}^{\rm e},-\bm{\eta}^{\rm o})
=
K_{ji}^{FU}(\bm{\eta}^{\rm e},\bm{\eta}^{\rm o}).
\end{align}
We emphasize that the Onsager-Casimir reciprocity does not, in general, imply that the resistance tensor is symmetric for a given fluid.
Rather, it relates the resistance tensor to its transpose in the auxiliary fluid, where $\bm\eta^{\rm o}$ changes its sign while $\bm\eta^{\rm e}$ remains unchanged. 
The conventional symmetry relation is recovered in the absence of odd viscosities.

Along similar lines, one can consider a rigid particle with surface velocity $\mathbf{v}=-\mathbf{E}^\infty\cdot\mathbf{r}$ and the same particle with a rotational motion $\mathbf{\hat{v}}=\bm{\hat{\Omega}}\times\mathbf{r}$.
Inserting these surface velocities into equation~\eqref{eq:LRT} yields
\begin{align}
    -E^\infty_{ij}\int_\mathcal{S} dS\,
	r_i 
    \hat{\sigma}_{jk}n_k
    =
    \hat{\Omega}_j
    \int_\mathcal{S} dS\,
    \epsilon_{ijk}r_k\sigma_{i\ell} n_\ell
    ,
\end{align}
where $ \epsilon_{ijk}$ is the Levi-Civita permutation symbol. 
Through the expressions of the torque and stresslet on the particle, one finds the scalar relation $-E^\infty_{ij}\hat{S}_{ij}=\hat{\Omega}_iM_i$, where we have used the fact that the tensors $\mathbf{E}^\infty$ and $\mathbf{S}$ are symmetric.
Using the definitions of the resistance tensors \eqref{eq:grand} gives $E^\infty_{ij}\hat{K}^{S\Omega}_{ijk}\hat{\Omega}_k=\hat{\Omega}_iK_{ijk}^{ME}E^\infty_{jk}$.
Sorting the right- and left-hand sides of the equation together as before, we find the relation using the conditions~\eqref{eq:cond}
\begin{align}
	K^{S\Omega}_{ijk} (\bm{\eta}^{\rm e},-\bm{\eta}^{\rm o})= 
	K^{ME}_{kij} (\bm{\eta}^{\rm e},\bm{\eta}^{\rm o})
	.
\end{align}

Applying similar arguments to the other resistance tensors and using the generalized reciprocal theorem, we conclude the following symmetries
\begin{align}
K_{ij}^{FU}(\bm{\eta}^{\rm e},-\bm{\eta}^{\rm o})
&= K_{ji}^{FU}(\bm{\eta}^{\rm e},\bm{\eta}^{\rm o}),
&
K_{ij}^{M\Omega}(\bm{\eta}^{\rm e},-\bm{\eta}^{\rm o})
&= K_{ji}^{M\Omega}(\bm{\eta}^{\rm e},\bm{\eta}^{\rm o}),
\notag\\
K_{ij}^{F\Omega}(\bm{\eta}^{\rm e},-\bm{\eta}^{\rm o})
&= K_{ji}^{MU}(\bm{\eta}^{\rm e},\bm{\eta}^{\rm o}),
&
K_{ijk}^{SU}(\bm{\eta}^{\rm e},-\bm{\eta}^{\rm o})
&= K_{kij}^{FE}(\bm{\eta}^{\rm e},\bm{\eta}^{\rm o}),
\label{eq:resistance}
\\
K_{ijk}^{S\Omega}(\bm{\eta}^{\rm e},-\bm{\eta}^{\rm o})
&= K_{kij}^{ME}(\bm{\eta}^{\rm e},\bm{\eta}^{\rm o}),
&
K_{ijk\ell}^{SE}(\bm{\eta}^{\rm e},-\bm{\eta}^{\rm o})
&= K_{k\ell ij}^{SE}(\bm{\eta}^{\rm e},\bm{\eta}^{\rm o}).
\notag
\end{align}
Remarkably, the obtained symmetry relations suggest that for an arbitrarily shaped particle, there are six tensorial blocks needed to determine the resistance on the particle even in a fluid characterized by a full viscosity tensor.

The above symmetries are manifestations of Onsager-Casimir reciprocity, which has been widely discussed in systems with broken time-reversal and parity symmetries \citep{fruchart2023odd, everts2024dissipative, hosaka2024chirotactic, khain2024trading}. 
Here, we have established that these relations hold for rigid bodies of arbitrary shape. Although the presence of odd viscosity generally renders the resistance matrix nonsymmetric, its invertibility can still be guaranteed by the positive definiteness of its symmetric part. Specifically, we assume that the dissipative part of the viscosity tensor is strictly positive definite, ensuring positive viscous dissipation for any nonzero rigid-body motion. Consequently, the symmetric part of the resistance matrix is positive definite, which guarantees the invertibility of the full resistance matrix, even in the presence of an antisymmetric contribution. 
This assumption excludes the limiting case of a purely odd-viscous fluid.

\section{Generalized swimming formula}
\label{sec:swimming}

Having discussed the allowed tensorial structures of the resistance tensors for generalized viscous fluids, we here extend this linear force moment-velocity relation to include the propulsive speed of swimming objects.
We assume that the main solution $(\mathbf{v},\bm{\sigma}=-p\mathbf{I}+\bm\eta:\mathbf{E})$ corresponds to solutions around a deforming swimmer (the deformation or surface slip velocity $\mathbf{v}^s$) whose surface velocity is given by 
\begin{align}
    \mathbf{v}&=\mathbf{U}+\bm{\Omega}\times\mathbf{r}+\mathbf{v}^s,
\end{align}
in an external flow $\mathbf{v}^\infty$ as in equation~\eqref{eq:infinity}, and that the auxiliary solution with the hat symbol $(\mathbf{\hat{v}},\bm{\hat{\sigma}}=-\hat{p}\mathbf{I}+\bm{\hat{\eta}}:\mathbf{\hat{E}})$ is chosen for a flow around a rigid body with arbitrary translational and rotational velocities in the absence of a background flow
\begin{align}
    \mathbf{\hat{v}}
    &=
    \mathbf{\hat{U}}+\bm{\hat{\Omega}}\times\mathbf{r}.
\end{align}
In the above, we have defined the rate of strain tensors, such that $\mathbf{E}=\frac{1}{2}[\nabla\mathbf{v}+(\nabla\mathbf{v})^\top]$ and $\mathbf{\hat{E}}=\frac{1}{2}[\nabla\mathbf{\hat{v}}+(\nabla\mathbf{\hat{v}})^\top]$ and the double dot products are defined such that $[\bm{\eta}:\mathbf{E}]_{ij}=\eta_{ijk\ell}E_{k\ell}$.

The generalized Lorentz reciprocal theorem evaluated at the particle surface follows from equation~\eqref{eq:LRT}
\begin{align}
    \int_\mathcal{S} dS\, 
    (
    \mathbf{v}
    -
    \mathbf{v}^\infty
    )
    \cdot
    \hat{\bm{\sigma}}
    \cdot
    \mathbf{n}
    =
    \int_\mathcal{S} dS\, 
    \mathbf{\hat{v}}
    \cdot
    (\bm{\sigma}-\bm{\sigma}^\infty)
    \cdot
    \mathbf{n}
    .
\end{align}
It is convenient to introduce the translational and rotational surface force resistance tensors, $\boldsymbol{\hat{\Sigma}}^U$ and $\boldsymbol{\hat{\Sigma}}^\Omega$, such that the surface force, $\hat{\mathbf{f}}=-\hat{\boldsymbol{\sigma}}\cdot\mathbf{n}$, can be decomposed as
\begin{align}
    \hat{\mathbf{f}}=
    -\boldsymbol{\hat{\Sigma}}^U\cdot\hat{\mathbf{U}}
    -\boldsymbol{\hat{\Sigma}}^\Omega\cdot\hat{\boldsymbol{\Omega}}
    ,
\end{align}
where we have used the linearity of the flow equation.
We emphasize that the translational and rotational surface force resistance tensors, $\hat{\boldsymbol{\Sigma}}^U$ and $\hat{\boldsymbol{\Sigma}}^\Omega$, are defined for the auxiliary problem, whose viscosity tensors satisfy $\hat{\boldsymbol{\eta}}^{\mathrm e}=\boldsymbol{\eta}^{\mathrm e}$ and $\hat{\boldsymbol{\eta}}^{\mathrm o}=-\boldsymbol{\eta}^{\mathrm o}$. 
In the absence of the odd viscosity tensor in both problems, they have identical constitutive relations in the reciprocal theorem, so that $\hat{\boldsymbol{\Sigma}}^U=\boldsymbol{\Sigma}^U$ and $\hat{\boldsymbol{\Sigma}}^\Omega=\boldsymbol{\Sigma}^\Omega$. 
This situation includes the classical Newtonian fluid with a shear viscosity alone \citep{ishimoto2023jeffery}, as well as fluids characterized exclusively by multiple even viscosities.

Using these surface force resistance tensors of the auxiliary problem and integrating them over the surface $\mathcal{S}$ gives the resistance tensors
\begin{align}
    \hat{K}_{ij}^{FU}
    &=
    \int_\mathcal{S}dS\,
    \hat{\Sigma}^U_{ij},
    \quad
    \hat{K}_{ij}^{F\Omega}
    =
    \int_\mathcal{S}dS\,
   \hat{\Sigma}_{ij}^\Omega,
 \end{align}
 for the linear force, and
 \begin{align}
    \hat{K}_{ij}^{MU}
    &=
    \int_\mathcal{S}dS\,
    \epsilon_{ik\ell}r_k
    \hat{\Sigma}^U_{\ell j},
    \quad
    \hat{K}_{ij}^{M\Omega}
    =
    \int_\mathcal{S}dS\,
    \epsilon_{ik\ell}r_k
    \hat{\Sigma}^\Omega_{\ell j}
       \label{eq:KMUMO}
    ,
\end{align}
for the torque moment.
These are integrated into the grand resistance tensor, a single $6\times6$ matrix
\begin{align}
\bm{\mathcal{\hat{K}}}= 
 \begin{pmatrix}
        \mathbf{\hat{K}}^{FU} & \mathbf{\hat{K}}^{F\Omega}\\
        \mathbf{\hat{K}}^{MU} & \mathbf{\hat{K}}^{M\Omega}
    \end{pmatrix},
    \label{eq:mathcalK}
\end{align}
which relates the hydrodynamic force and torque to the translational and angular velocities:
\begin{align}
    \begin{pmatrix}
    \mathbf{\hat{F}}\\
    \mathbf{\hat{M}}
    \end{pmatrix}
    =
    -
    \bm{\hat{\mathcal{K}}}
   \begin{pmatrix}
    \mathbf{\hat{U}}\\
    \boldsymbol{\hat{\Omega}}
    \end{pmatrix}
    .
\end{align}

Using the decomposition of the traction $\mathbf{\hat{f}}$ as before, the reciprocal theorem can be written as 
\begin{align}
    \mathbf{U}\cdot\mathbf{\hat{F}}
    +
    \bm{\Omega}\cdot\mathbf{\hat{M}}
    +
    \int_\mathcal{S} dS\, 
    (\mathbf{v}^s
    -
    \mathbf{v}^\infty)\cdot
    [-\boldsymbol{\hat{\Sigma}}^U\cdot\hat{\mathbf{U}}
    -\boldsymbol{\hat{\Sigma}}^\Omega\cdot\hat{\boldsymbol{\Omega}}]
    =
    \mathbf{\hat{U}}\cdot\mathbf{F}
    +
    \bm{\hat{\Omega}}\cdot\mathbf{M}
    .
    \label{eq:UFOM}
\end{align}
For a spatially uniform, symmetric background stress, the net force and torque contributions vanish.
Defining the force moments due to the surface activity and the background flow, such that 
\begin{align}
    \mathbf{F}^{prop} = -\int_\mathcal{S}dS\,
    \mathbf{v}^s\cdot\boldsymbol{\hat{\Sigma}}^U,
    \quad
    \mathbf{M}^{prop} = -\int_\mathcal{S}dS\,
    \mathbf{v}^s\cdot\boldsymbol{\hat{\Sigma}}^\Omega,
    \label{eq:UMprop}
\end{align}
and
\begin{align}
    \mathbf{F}^\infty = 
    \int_\mathcal{S}dS\,
    \mathbf{v}^\infty\cdot\boldsymbol{\hat{\Sigma}}^U,
    \quad
    \mathbf{M}^\infty = 
    \int_\mathcal{S}dS\,
    \mathbf{v}^\infty\cdot\boldsymbol{\hat{\Sigma}}^\Omega,
    \label{eq:Finf}
\end{align}
we find the following integral formula
\begin{align}
    \mathbf{U}\cdot\hat{\mathbf{F}}
    +
    \boldsymbol{\Omega}\cdot\hat{\mathbf{M}}
    +
    (\mathbf{F}^{prop}+\mathbf{F}^\infty)\cdot\hat{\mathbf{U}}
    +
    (\mathbf{M}^{prop}+ \mathbf{M}^\infty)\cdot\hat{\boldsymbol{\Omega}}
    =
    \hat{\mathbf{U}}\cdot\mathbf{F}
    +
    \hat{\boldsymbol{\Omega}}\cdot\mathbf{M}
    .
    \label{eq:integralKI}
\end{align}
Noting that $\hat{\mathbf{U}}$ and $\hat{\boldsymbol{\Omega}}$ can be taken arbitrarily, equation~\eqref{eq:integralKI} may further be reduced to
\begin{align}
    \begin{pmatrix}
    \mathbf{F}\\
    \mathbf{M}
    \end{pmatrix}
    =
    -
    \bm{\mathcal{\hat{K}}}^\top
    \begin{pmatrix}
    \mathbf{U} \\
    \boldsymbol{\Omega} \\
    \end{pmatrix}
    +
    \begin{pmatrix}
    \mathbf{F}^\infty \\
    \mathbf{M}^\infty \\
    \end{pmatrix}
    +
    \begin{pmatrix}
    \mathbf{F}^{prop}\\
    \mathbf{M}^{prop}\\
    \end{pmatrix}
    \label{eq:UOmega}
    .
\end{align}

Expression~\eqref{eq:UOmega} can be expressed with the swimming velocities and the background flow velocities for a swimmer under the force- and torque-free conditions.
Inserting the linear background flow velocity in equation~\eqref{eq:infinity} into equation~\eqref{eq:Finf}, one finds the flow-induced force and torque as
\begin{align} 
F_{i}^{\infty} & = 
U_{j}^{\infty} \hat{K}_{ji}^{F U} 
+\Omega_{j}^{\infty} \hat{K}_{ji}^{MU} 
+ \hat{K}_{ijk}^{FE}E_{jk}^{\infty},
\\ 
M_{i}^{\infty} 
& 
=U_{j}^{\infty} \hat{K}_{ji}^{F\Omega} 
+ \Omega_{j}^{\infty} \hat{K}_{ji}^{M \Omega} 
+  \hat{K}_{ijk}^{ME}E_{jk}^{\infty},
\end{align}
where the third-rank tensors $\hat{K}_{i j k}^{FE}$ and $\hat{K}^{ME}_{i j k}$ are, respectively, called the force-shear tensor and torque-shear tensor, defined as
\begin{align}
    \hat{K}_{i j k}^{FE}=\frac{1}{2} \int_{S} dS\, \left( r_k \hat{\Sigma}_{j i}^{U}+r_j \hat{\Sigma}_{k i}^{U} \right), \quad 
    \hat{K}^{ME}_{i j k}=\frac{1}{2} \int_{S} dS\, \left( r_k \hat{\Sigma}_{j i}^{\Omega}+r_j \hat{\Sigma}_{k i}^{\Omega} \right).
\end{align}
With the symmetric property of the rate-of-strain tensor,
these third-rank tensors are symmetric under the exchanges of the last two indices, namely, $\hat{K}_{ijk}^{FE}=\hat{K}_{ikj}^{FE}$ and $\hat{K}_{ijk}^{ME}=\hat{K}_{ikj}^{ME}$.
When $\mathbf{F}=\mathbf{M}=\mathbf{0}$, equation~\eqref{eq:UOmega} can then be reduced to the swimming formula that links the velocities of a free swimmer of arbitrary shape with its prescribed surface-actuation velocity
\begin{align}
    \begin{pmatrix}
    \mathbf{U} \\
    \boldsymbol{\Omega}
    \end{pmatrix}
    =
    \begin{pmatrix}
    \mathbf{U}^\infty \\
    \boldsymbol{\Omega}^\infty
     \end{pmatrix}
    +
    (\bm{\hat{\mathcal{K}}}^\top)^{-1}
    \begin{pmatrix}
    \mathbf{\hat{K}}^{FE}:\mathbf{E}^\infty+    \mathbf{F}^{prop}\\
     \mathbf{\hat{K}}^{ME}:\mathbf{E}^\infty+    \mathbf{M}^{prop}\\
    \end{pmatrix}
    \label{eq:UOmega2}
    .
\end{align}
We retain the hats on quantities associated with the auxiliary problem to emphasize that the auxiliary fluid is characterized by the odd viscosity tensor $-\bm{\eta}^{\rm o}$ rather than $\bm{\eta}^{\rm o}$, as required by the reciprocal theorem when condition \eqref{eq:cond} holds.

We now discuss several implications of the derived swimming formula generalized to a linearly viscous fluid characterized by a viscosity tensor.
First, the grand resistance tensor $\bm{\mathcal{\hat{K}}}$ and the shear-induced resistance tensors $\mathbf{\hat{K}}^{FE}$ and $\mathbf{\hat{K}}^{ME}$ are characterized by a viscosity tensor, which can contain both even and odd components.
Nevertheless, the generalized swimming formula retains the same algebraic structure as the classical one: the translational and rotational velocities are related to the propulsive and shear-induced force moments through the inverse of the grand resistance matrix \citep{stone1996propulsion, yariv2006self, ishimoto2023jeffery}.
In the Newtonian limit, the conventional swimming formula is recovered.
Second, despite this formal similarity, the propulsive and shear-induced velocities involve the surface force resistance tensors of the auxiliary problem.
The main and auxiliary problems therefore involve opposite signs of the odd viscosity tensor, as required by the generalized Lorentz reciprocal theorem.
This distinction requires careful treatment when applying the reciprocal theorem to particle and swimmer dynamics, as demonstrated in previous studies considering a single odd viscosity coefficient \citep{hosaka2023lorentz, hosaka2024chirotactic, chatzittofi2026self, lier2024odd, lier2026sorting, van2026linear}.

\section{Scallop theorem in fluids governed by a generalized linear constitutive relation}
\label{sec:scallop}

We have derived the swimming formula for generalized linearly viscous fluids.
To demonstrate its applicability, we employ it to provide a proof of the scallop theorem, which states that a microswimmer undergoing time-reversible (reciprocal) body deformations cannot achieve net displacement in a Stokes flow.
Here we extend this classical theorem formulated in a low-Reynolds number Newtonian fluid to fluids governed by the generalized linear constitutive relation.
Although swimming dynamics in chiral fluids have been studied before \citep{lapa2014, hosaka2023hydrodynamics, hosaka2023lorentz, hosaka2024chirotactic, chatzittofi2026self}, a general proof of the scallop theorem for the class of fluids considered here is still needed.

We follow a coordinate-based proof of the scallop theorem \citep{ishimoto2012coordinate}, which considers a microswimmer and a virtual swimmer in the absence of an external flow.
The generalized swimming formula \eqref{eq:UOmega2} confirms that the rigid-body translational and rotational velocities in a quiescent fluid depend linearly on the propulsive force and torque moments through the inverse transpose of the auxiliary grand resistance matrix.
Although the resistance operator depends on the instantaneous shape and orientation, the rigid-body velocity is linear in the prescribed deformation rate. 
For a reciprocal deformation cycle, reversing the shape path reverses the complete configuration-space trajectory, provided that the instantaneous boundary-value problem has a unique solution.
In this sense, since $\mathbf{F}^{prop}$ is itself a linear functional of the instantaneous surface velocity $\mathbf{v}^s$ [see equation~\eqref{eq:UMprop}], the entire kinematic relationship is still linear, despite the nontrivial combinations of the even and odd viscosity tensors, $\bm{\eta}^{\rm e}$ and $\bm{\eta}^{\rm o}$.
If the sequence of deformations is reciprocal, the swimmer merely retraces its shape variations, $\mathbf{v}^s$, backward in time during the second half of the cycle.
Integrating the linear velocity over a complete time-reversible cycle therefore identically yields zero net displacement, demonstrated in the same velocity-surface deformation formalism given in \citep{ishimoto2012coordinate}.

\begin{figure}[htbp]
    \centering
    \includegraphics[width=0.6\textwidth]{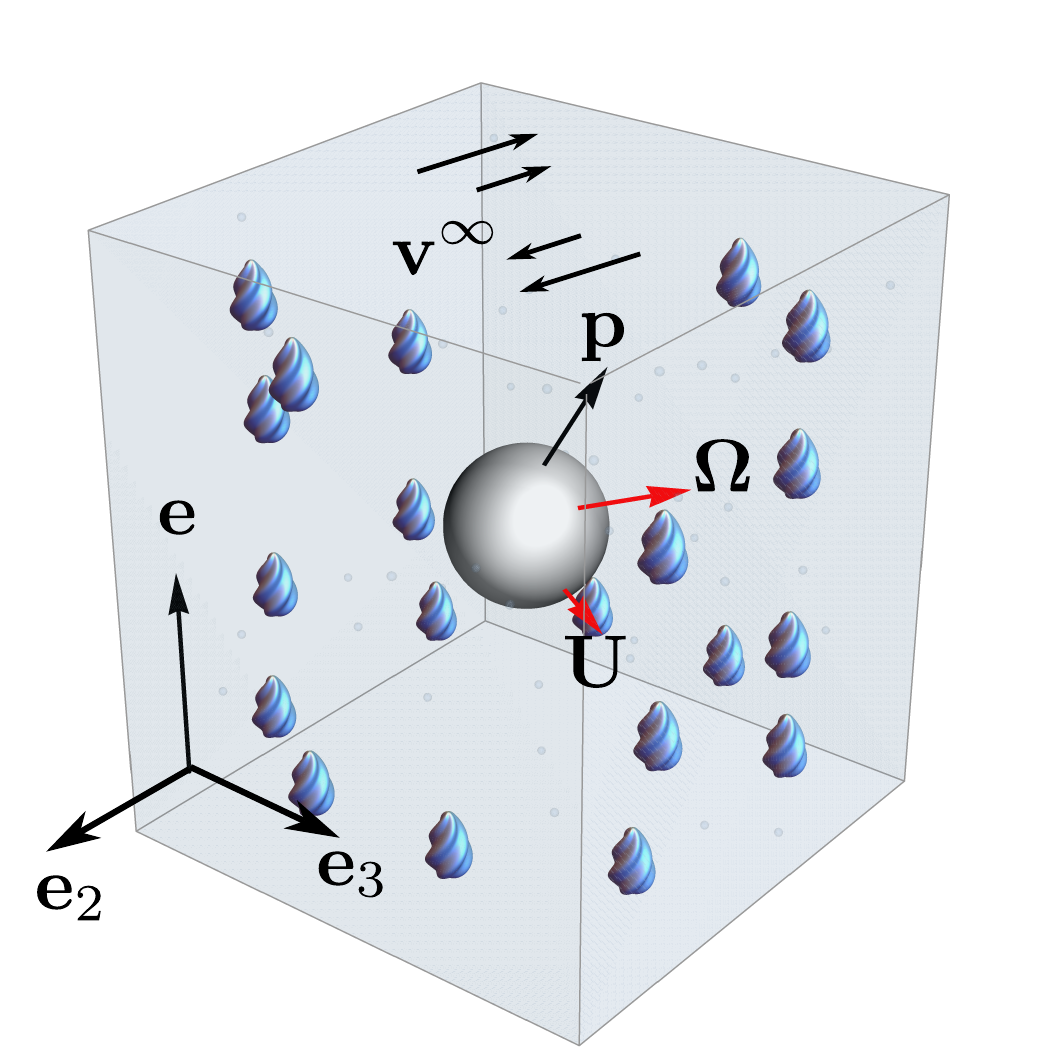}
    \caption{
Schematic illustration of a spherical swimmer under a linear shear flow in a chiral uniaxial fluid with the axis of symmetry $\mathbf{e}$.
We consider the laboratory-fixed reference frame $\{\mathbf{e},\mathbf{e}_2,\mathbf{e}_3\}$, and the body-fixed director $\mathbf{p}$.
The linear and angular velocities of the object are $\mathbf{U}$ and $\boldsymbol{\Omega}$, respectively. 
The background linear flow field is denoted as $\mathbf{v}^\infty$.
The suspended particles (blue) aligned along the chiral axis $\mathbf{e}$ represent the chiral uniaxial nature of the effective surrounding fluid (see figure~\ref{fig:symmeties} for different symmetry classes).
    }
    \label{fig:sphere}
\end{figure}

\section{Spherical body in a chiral uniaxial fluid}
\label{sec:resistance}

\subsection{Resistance tensors}

In a Newtonian fluid, the tensorial structure of the hydrodynamic resistance of an object is constrained by its geometry, since the fluid itself introduces no preferred direction. Here, we consider a sphere immersed in a chiral uniaxial fluid characterized by a single distinguished axis about which the fluid is invariant under rotations (figure \ref{fig:sphere}). 
Fluids with such axial symmetry have been studied in the context of transversely isotropic fluids \citep{ericksen1960transversely}. Here, we extend this framework to incorporate chiral symmetry and both even and odd contributions to the viscosity tensor.
Symmetry-based arguments have previously been employed to determine the hydrodynamic response of helicoidal particles in shear flows \citep{brenner1964stokesa, brenner1964stokesb, ishimoto2020helicoidal,ishimoto2020jeffery}. 
Here, we generalize this approach from the symmetry of a suspended particle to the symmetry of the constitutive relation of the surrounding fluid.
In particular, by considering a spherical particle whose geometry is isotropic, we can directly identify how the fluid symmetry constrains the tensorial structure of its hydrodynamic resistance \citep{khain2024trading}. 
The continuous rotational symmetry about the fluid axis reduces the number of independent coefficients appearing in the six resistance tensors relevant to the particle's dynamics.

In the following, we investigate the resulting second- and third-rank resistance tensors, relevant to the particle dynamics, and infer their tensorial structures from the fluid that satisfies the continuous rotational symmetry.
The symmetry of a chiral uniaxial fluid is illustrated in figure \ref{fig:symmeties}(a), while the cases with additional symmetries are summarized in figure \ref{fig:symmeties}(b-d).

\subsection{Second-rank resistance tensors}

We first consider a second-rank resistance tensor satisfying the continuous rotational symmetry around a unit vector $\mathbf{e}_1$, where we employ a right-handed orthogonal basis $\{\mathbf{e}_1,\mathbf{e}_2,\mathbf{e}_3\}$ in the laboratory frame.
Under this constraint, the diagonal sub-matrices are allowed to take the following forms, each of which is characterized by three coefficients \citep{khain2024trading}
\begin{align}
    \mathbf{\hat{K}}^{FU}
    &= 
    K^{FU}_1\mathbf{ee}
    +
    K^{FU}_2 (\mathbf{I}-\mathbf{ee})
    +
    K^{FU}_{23}
    (\mathbf{e}_2\mathbf{e}_3-\mathbf{e}_3\mathbf{e}_2),
    \label{eq:KFU}
    \\
    \mathbf{\hat{K}}^{M\Omega}
    &= 
    K^{M\Omega}_1\mathbf{ee}
    +
    K^{M\Omega}_2 (\mathbf{I}-\mathbf{ee})
    +
    K^{M\Omega}_{23}
    (\mathbf{e}_2\mathbf{e}_3-\mathbf{e}_3\mathbf{e}_2),
    \label{eq:KMO}
\end{align}
where we have redefined $\mathbf{e}_1\equiv\mathbf{e}$ for brevity, and the hats on the scalar coefficients are omitted for notational simplicity, while those on the resistance tensors are retained to indicate that they are from the auxiliary problem in the swimming formula in \eqref{eq:UOmega2}.
Note here that the last terms in the above can also be expressed using the third-rank Levi-Civita tensor $\bm{\epsilon}$, such that $\bm{\epsilon}\cdot\mathbf{e}=\mathbf{e}_2\mathbf{e}_3-\mathbf{e}_3\mathbf{e}_2$, as used in previous studies on odd viscosity \citep{hosaka2023lorentz, khain2024trading}.
The scalars $K_1^{FU},K_2^{FU},K_1^{M\Omega},$ and $K_2^{M\Omega}$ are all positive, while the two chiral coefficients $K^{FU}_{23}$ and $K^{M\Omega}_{23}$ can take either positive or negative values, as required from the Onsager-Casimir reciprocity.

Similarly, the resistance tensors that couple translational and rotational modes can be written as 
\begin{align}
    \mathbf{\hat{K}}^{F\Omega}
    &=
    K^{F\Omega}_1\mathbf{ee}
    +
    K^{F\Omega}_2(\mathbf{I}-\mathbf{ee})
    +
    C_{23}
    (\mathbf{e}_2\mathbf{e}_3-\mathbf{e}_3\mathbf{e}_2)
    ,
    \label{eq:KFO}
   	\\
   \mathbf{\hat{K}}^{MU}
   &=
    {K}^{MU}_1\mathbf{ee}
    +
    {K}^{MU}_2(\mathbf{I}-\mathbf{ee})
    -
    C_{23}
    (\mathbf{e}_2\mathbf{e}_3-\mathbf{e}_3\mathbf{e}_2)
        \label{eq:KMU}
    .
\end{align}
Under the mirror reflection transformation across the plane containing the symmetry axis, i.e., $\mathbf{e}_2\to-\mathbf{e}_2$, the resistance coefficients appearing in the first and second terms vanish $(K^{F\Omega}_1=K^{F\Omega}_2=K^{MU}_1=K^{MU}_2=0)$, meaning that these coefficients result from the fluid chirality, while $C_{23}$ is parity-even.
Importantly, the chiral coefficients $K^{F\Omega}_1,K^{F\Omega}_2,K^{MU}_1,$ and $K^{MU}_2$ can further be decomposed into the two contributions: one purely from even components and the other due to odd components under the transformation $\bm{\eta}^{\rm o}\to-\bm{\eta}^{\rm o}$.
These two effects can be distinguished as follows.
The contributions from the even parts do not change its sign under the permutation $i\leftrightarrow j$ of the coupling tensors $\mathbf{\hat{K}}^{F\Omega}$ and $\mathbf{\hat{K}}^{MU}$, while those flip sign following the Onsager-Casimir reciprocity in equation~\eqref{eq:resistance}.
Consequently, in the presence of odd terms, $K_1^{F\Omega}\neq K_1^{MU}$ and $K_2^{F\Omega}\neq K_2^{MU}$, while the Onsager-Casimir reciprocity relates each coefficient to its counterpart in the fluid with the signs of the odd viscosity coefficients reversed.
The parity-even terms with $C_{23}$ can be eliminated by choosing the origin of the body-fixed frame taken to coincide with the hydrodynamic center of resistance \citep{khain2024trading}.

\subsection{Third-rank resistance tensors}

A third-rank tensor compatible with the continuous rotational symmetry can be expressed in terms of three coefficients \citep{ishimoto2020jeffery}.
Specifically, the force-shear and torque-shear tensors take the forms of 
\begin{align}
    \hat{K}^{FE}_{ijk}
    &=
    K^{FE}_1
   e_ie_je_k
    +
    \frac{1}{2}
    K^{FE}_2
    \left[
    (
    \delta_{ij}-e_ie_j
    )
    e_k
    +
    (
    \delta_{ik}-e_ie_k
    )
    e_j
    \right]
    \nonumber\\
    &
    +
    \frac{1}{2}
    K^{FE}_{23}
    [
    (
    e_{2,i}e_{3,j}
    -
    e_{3,i}e_{2,j}
    )
    e_k
    +
    (
    e_{2,i}e_{3,k}
    -
    e_{3,i}e_{2,k}
    )
    e_j
    ]
    ,
        \label{eq:KFE}
    \\
    \hat{K}_{ijk}^{ME}
    &=
    K^{ME}_1
   e_ie_je_k
    +
    \frac{1}{2}
    K^{ME}_2
    \left[
    (
    \delta_{ij}-e_ie_j
    )
    e_k
    +
    (
    \delta_{ik}-e_ie_k
    )
    e_j
    \right]
    \nonumber\\
    &
    +
    \frac{1}{2}
    K^{ME}_{23}
    [
    (
    e_{2,i}e_{3,j}
    -
    e_{3,i}e_{2,j}
    )
    e_k
    +
    (
    e_{2,i}e_{3,k}
    -
    e_{3,i}e_{2,k}
    )
    e_j
    ]
    .
    \label{eq:KME}
\end{align}
Notice that the $1/2$ prefactors of the terms involving $K_2^{FE},K_{23}^{FE}, K_2^{ME},$ and $K_{23}^{ME}$ follow the convention used in the previous study about helicoidal objects \citep{ishimoto2020helicoidal}.
Further symmetry arguments identify which terms result from odd effects.
Components that are invariant under reflection containing the $\mathbf{e}$-axis are parity-even, while those that change their sign are parity-odd, as before.
Noting that the force transforms as $(F_1,F_2,F_3)\to(F_1,-F_2,F_3)$ in the orthogonal basis $\{\mathbf{e},\mathbf{e}_2,\mathbf{e}_3\}$ under the mirror reflection $\mathbf{e}_2\to-\mathbf{e}_2$, the coefficients $K_1^{FE}$ and $K_2^{FE}$ are invariant under this transformation, while $K_{23}^{FE}$ is not invariant.
Hence, $K_1^{FE}$ and $K_2^{FE}$ are parity-even and $K_{23}^{FE}$ is parity-odd.
Given the torque transforms as $(M_1,M_2,M_3)\to(-M_1,M_2,-M_3)$, one finds $K_1^{ME},K_2^{ME}$ are parity-odd, while $K_{23}^{ME}$ is parity-even.
In principle, the parity-odd coefficients $K^{FE}_{23},K^{ME}_{1},K^{ME}_{2}$ can include terms resulting from the odd components in the viscosity tensor, akin to $K_1^{F\Omega},K_2^{F\Omega},K_1^{MU},K_2^{MU}$ in the coupling tensors.

\section{Dynamics of a spherical body in a chiral uniaxial fluid under a shear flow}
\label{sec:dynamics}

Here we investigate the translational and rotational velocities of a spherical object immersed in a chiral uniaxial fluid.
To derive the velocities, we start with the swimming formula derived in \eqref{eq:UOmega2}.
One can obtain the swimming formula by specifying the grand resistance tensor, $\bm{\mathcal{\hat{K}}}$, and the shear-induced force and torque, $\mathbf{\hat{K}}^{FE}$ and $\mathbf{\hat{K}}^{ME}$.
For later convenience, we give the expression of the inverse of $\bm{\mathcal{\hat{K}}}$ (equivalently, the mobility tensor), which is defined as
\begin{align}
    \bm{\mathcal{\hat{K}}}
    ^{-1}
    =
    \begin{pmatrix}
        \bm{\mathcal{M}}_{11} & \bm{\mathcal{M}}_{12}\\
        \bm{\mathcal{M}}_{12}^\ast &
        \bm{\mathcal{M}}_{22} 
    \end{pmatrix}
    ,
\end{align}
where the symbol $\ast$ denotes the transposition accompanied by the sign reversal of the resistance coefficients associated with the odd viscosity tensor, while keeping the even viscosity tensor unchanged.
Inserting the resistance tensors of a sphere in a chiral uniaxial fluid in equations \eqref{eq:KFU}-\eqref{eq:KMU} into $\bm{\mathcal{\hat{K}}}^{-1}$ yields the submatrices of the mobility tensor
\begin{align}
    \bm{\mathcal{M}}_{11} 
    &= 
    \frac{K^{M\Omega}_1}{\Delta_1} \mathbf{ee}
    +
    \frac{
    K^{M\Omega}_2L+K^{M\Omega}_{23}T}{\Delta_2}
    (\mathbf{I-ee})
    +
    \frac{
    K^{M\Omega}_{23} L-K^{M\Omega}_2T}
    {\Delta_2}
    (\mathbf{e}_2\mathbf{e}_3-\mathbf{e}_3\mathbf{e}_2)
    ,\\
    \bm{\mathcal{M}}_{12} 
    &= 
    -\frac{K^{F\Omega}_1}{\Delta_1}
    \mathbf{ee}
    -
    \frac{
    K^{F\Omega}_2L+C_{23}T}{\Delta_2}
    (\mathbf{I-ee})
    +
    \frac{
    K^{F\Omega}_{2}T -C_{23}L}
    {\Delta_2}
    (\mathbf{e}_2\mathbf{e}_3-\mathbf{e}_3\mathbf{e}_2)
    ,\\
    \bm{\mathcal{M}}_{22} 
    &= 
    \frac{K^{FU}_1}{\Delta_1} \mathbf{ee}
    +
    \frac{
    K^{FU}_2L+K^{FU}_{23}T}{\Delta_2}
    (\mathbf{I-ee})
    +
    \frac{
    K^{FU}_{23}L -K^{FU}_{2}T}
    {\Delta_2}
    (\mathbf{e}_2\mathbf{e}_3-\mathbf{e}_3\mathbf{e}_2)
    ,
\end{align}
where the parity-even determinants of the block matrix are given by
\begin{align}
    \Delta_1 &= 
    K^{FU}_1 K^{M\Omega}_1 
    -K^{F\Omega}_{1}K^{MU}_{1}
    ,\quad
    \Delta_2
    =L^2+T^2,
\end{align}
with the scalar coefficients expressed with the quadratic forms of the components of the resistance tensors
\begin{align}
    L &=
    K^{FU}_2K^{M\Omega}_2-K^{FU}_{23}K^{M\Omega}_{23}
    -K^{F\Omega}_{2}K^{MU}_{2}-C_{23}^2,
\end{align}
for the parity-even term, and 
\begin{align}
    T &=
	K^{FU}_2K^{M\Omega}_{23}+K^{FU}_{23}K^{M\Omega}_{2}
   +C_{23}(K^{F\Omega}_2-K^{MU}_2).
\end{align}
for the parity-odd term that vanishes for fluids with even viscosities alone, i.e., $K_{23}^{FU}=K_{23}^{M\Omega}=0$ and $K_2^{F\Omega}=K_2^{MU}$.

Inserting the transpose of $\bm{\hat{\mathcal{K}}}^{-1}$ and $\mathbf{\hat{K}}^{FE}$ in \eqref{eq:KFE} and $\mathbf{\hat{K}}^{ME}$ in \eqref{eq:KME} into \eqref{eq:UOmega2} yields the linear and angular velocities of a spherical body in a chiral uniaxial fluid
\begin{align}
    \mathbf{U} &=
    \mathbf{U}^\infty+
    [
    \alpha_U \mathbf{ee}
    +
    \beta_U (\mathbf{I-ee})
    +
    \gamma_U
    (\mathbf{e}_2\mathbf{e}_3-\mathbf{e}_3\mathbf{e}_2)
    ]\cdot\mathbf{E}^\infty\cdot\mathbf{e}
    +\mathbf{U}^{prop},
        \label{eq:u_sym}\\
    \bm\Omega &=
    \bm{\Omega}^\infty+
    [
    \alpha_\Omega  \mathbf{ee}
    +
    \beta_\Omega (\mathbf{I-ee})
    +
    \gamma_\Omega
    (\mathbf{e}_2\mathbf{e}_3-\mathbf{e}_3\mathbf{e}_2)
    ]\cdot\mathbf{E}^\infty\cdot\mathbf{e}
    +
    \bm{\Omega}^{prop},
    \label{eq:omega_sym}
\end{align}
where the coefficients accounting for the components of the external shear $\mathbf{E}^\infty$ are given by 
\begin{align}
    \alpha_U
    &=
    \frac{K^{FE}_1K^{M\Omega}_1-K^{MU}_1K^{ME}_1}{\Delta_1}
    ,
    \quad
    \beta_U
    =
    \frac{L a_2-Ta_1}{\Delta_2}
    ,
    \quad
    \gamma_U
    =
    \frac{L a_1+Ta_2}{\Delta_2}
    ,\\
    \alpha_\Omega
    &=
    \frac{K^{FU}_1K^{ME}_1-K^{F\Omega}_1K^{FE}_1}{\Delta_1}
    ,
    \quad
	\beta_\Omega
    =
   \frac{L b_2-Tb_1}{\Delta_2},\quad
 \gamma_\Omega
    =
   \frac{L b_1+Tb_2}{\Delta_2},
\end{align}
with 
\begin{align}
    a_1
    &=
    -K^{M\Omega}_{23}K^{FE}_2
    +
    K^{M\Omega}_{2}K^{FE}_{23}
    -
    K^{MU}_2 K^{ME}_{23}
    -
    C_{23} K^{ME}_2
	,\\
    a_2
    &=
    K^{M\Omega}_{2}K^{FE}_2
    +
    K^{M\Omega}_{23}K^{FE}_{23}
    -
    K^{MU}_2 K^{ME}_{2}
    +
    C_{23} K^{ME}_{23}
    ,\\
    b_1
    &=
    -K^{F\Omega}_{2}K^{FE}_{23}
    -
    K^{FU}_{23}K^{ME}_{2}
    +
    K^{FU}_2 K^{ME}_{23}
    +
    C_{23} K^{FE}_2
    ,\\
    b_2
    &=
    -K^{F\Omega}_{2}K^{FE}_{2}
    +
    K^{FU}_{2}K^{ME}_{2}
    +
    K^{FU}_{23} K^{ME}_{23}
    -
    C_{23} K^{FE}_{23}
    ,
\end{align}
for the coefficients of the shear modes and
\begin{align}
    \mathbf{U}^{prop}
    &=
    \bm{\mathcal{M}}_{11}^\top
    \cdot\mathbf{F}^{prop}
    +
    (\bm{\mathcal{M}}_{12}^\ast)^\top
    \cdot\mathbf{M}^{prop},\\
    \bm{\Omega}^{prop}
    &=
    \bm{\mathcal{M}}_{12}^\top
        \cdot\mathbf{F}^{prop}
    +
    \bm{\mathcal{M}}_{22}^\top
    \cdot\mathbf{M}^{prop},
\end{align}
for the propulsive active force moments.
In a Newtonian fluid, the shear-induced velocities vanish, leaving only $\mathbf{U}^\infty$ and $\bm{\Omega}^\infty$ and the swimming contributions, $\mathbf{U}^{prop}$ and $\bm{\Omega}^{prop}$.
The nonzero shear-induced modes highlight the significance of a structured fluid, which is otherwise possible for an anisotropic object in a Newtonian medium.

There exist three distinctive velocity modes along the anisotropic axis $\mathbf{e}$ and along its perpendicular directions.
In a Newtonian fluid, the spherical body just experiences the motion due to the linear and angular background flows.
However, in uniaxial fluids with continuous rotational symmetries, shear also induces motion.
If we decompose the shear vector $\mathbf{E}^\infty\cdot\mathbf{e}$, we find from equations~\eqref{eq:u_sym} and \eqref{eq:omega_sym}
\begin{align}
    \mathbf{U} &=
    \mathbf{U}^\infty
    +
    \alpha_U q_\|\mathbf{q}
    +
    \beta_U 
    \mathbf{q}_2
    +
    \gamma_U
    \mathbf{q}_3
    +\mathbf{U}^{prop},\\
    \bm\Omega &=
    \bm{\Omega}^\infty
    +
    \alpha_\Omega q_\|\mathbf{q}
    +
    \beta_\Omega
    \mathbf{q}_2
    +
    \gamma_\Omega
    \mathbf{q}_3
    +\bm{\Omega}^{prop},
\end{align}
with $q_\|= \mathbf{e}\cdot\mathbf{E}^\infty\cdot\mathbf{e}$ and the newly defined orthogonal directions:
\begin{align}
    \mathbf{q}=\mathbf{q}_1
    =\mathbf{e},
    \quad
    \mathbf{q}_2=(\mathbf{I-ee})\cdot\mathbf{E}^\infty\cdot\mathbf{e}
    ,\quad
    \mathbf{q}_3 = 
    (\mathbf{e}_2\mathbf{e}_3-\mathbf{e}_3\mathbf{e}_2)
    \cdot\mathbf{E}^\infty\cdot\mathbf{e}
    ,
\end{align}
with $|\mathbf{q}_2|=|\mathbf{q}_3|$ from the symmetry argument.
We observe that the particle undergoes translational and rotational velocities along the three mutually independent directions set by the fluid anisotropy and chirality.
Notice that shear flows purely in a plane perpendicular to the axis of symmetry do not induce any additional motion relative to the background flow, as $\mathbf{E}^\infty\cdot\mathbf{e}=\mathbf{0}$.
Thus, out-of-plane shear modes are essential to investigate the effect of chirality in anisotropic media.

\begin{figure}[htbp]
    \centering
    \includegraphics[width=0.7\textwidth]{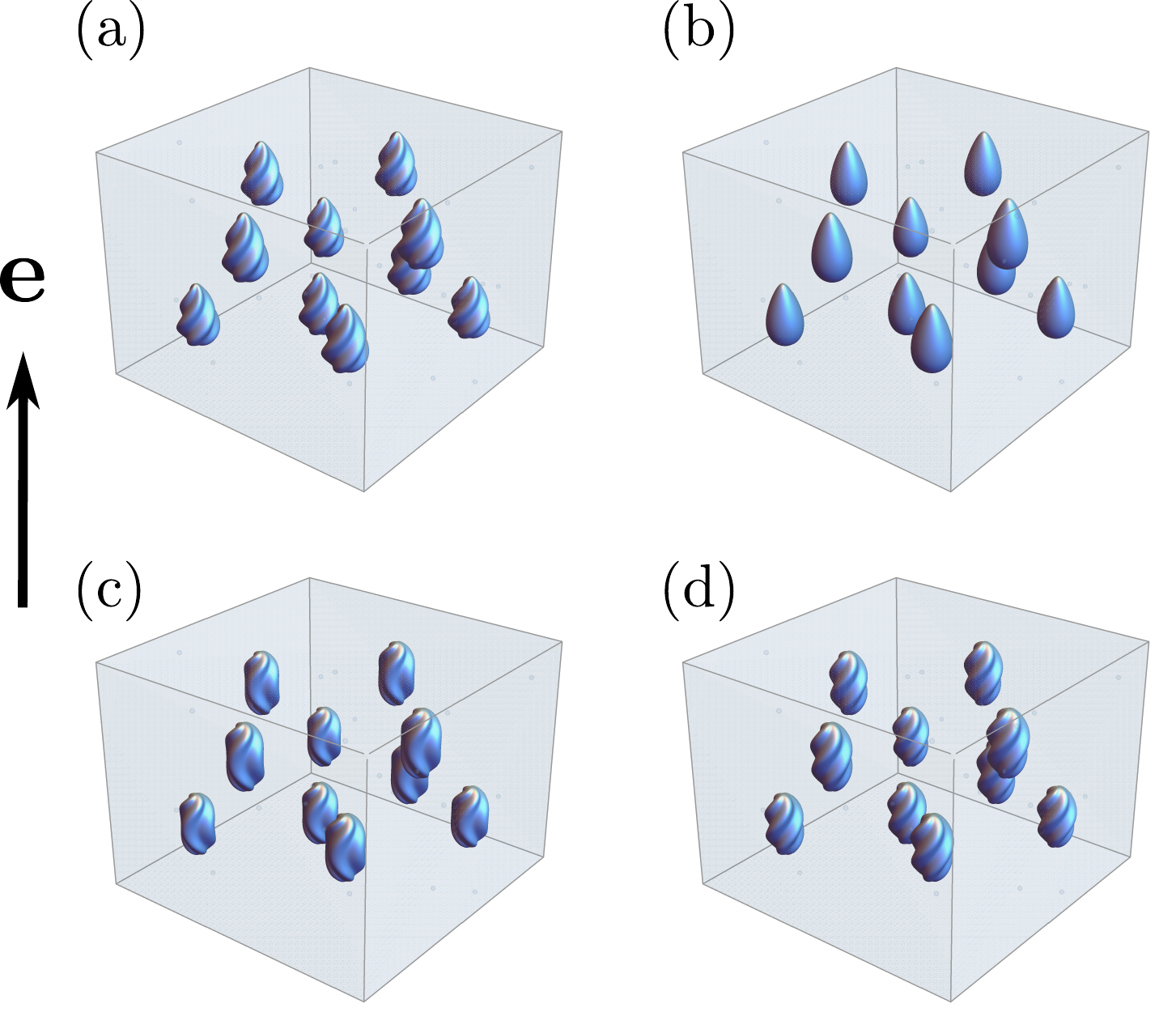}
    \caption{
Schematic illustrations of uniaxial fluids with different symmetry classes. 
The suspended particles serve as geometric illustrations of the corresponding symmetry classes; these fluid symmetries constrain the tensorial structures of resistance tensors.
For illustration purposes, the suspended objects in panels (a,c,d) exhibit $n$-fold symmetry about $\mathbf{e}$.
Although these objects possess the discrete four-fold rotational symmetry, this symmetry imposes the same constraints as continuous axial rotational symmetry on resistance tensors of rank up to three \citep{ishimoto2020jeffery}.
(a) A chiral uniaxial fluid of chiral bodies with $C_{4}$ symmetry about the axis $\mathbf{e}$, as considered in section~\ref{sec:resistance}.
(b) A transversely isotropic fluid of polar axisymmetric bodies with $C_{\infty v}$ symmetry about $\mathbf{e}$, as considered in section~\ref{sec:symmetry1}.
(c) A heterochiral fluid of bodies with $C_{4h}$ symmetry about $\mathbf{e}$, possessing reflection symmetry in a plane perpendicular to the axis, as considered in section~\ref{sec:symmetry2}.
(d) A homochiral fluid of bodies with $D_4$ symmetry, possessing two-fold rotational symmetry about axes perpendicular to $\mathbf{e}$, as considered in section~\ref{sec:symmetry3}.
    }
    \label{fig:symmeties}
\end{figure}

\subsection{Transversely isotropic fluid}
\label{sec:symmetry1}

With the additional symmetry constraint of reflectional symmetry in a plane containing the chiral axis, the fluid becomes achiral, as schematically shown in figure~\ref{fig:symmeties}(b).
Such a uniaxial system with this class of symmetry is known as a transversely isotropic fluid \citep{ericksen1960transversely}.
In this case, we have $K^{FU}_{23}=K^{M\Omega}_{23}=K^{F\Omega}_{1}=K^{F\Omega}_{2}=K^{MU}_{1}=K^{MU}_{2}=K^{FE}_{23}=K^{ME}_1=K^{ME}_2=0$.
The coefficients in the translational and angular velocities then reduce to
\begin{align}
    \alpha_U =\frac{K^{FE}_1}{K^{FU}_1}, \quad
    \beta_U  = 
    \frac{K^{M\Omega}_2K^{FE}_{2}+C_{23}K^{ME}_{23}}{K^{FU}_2K^{M\Omega}_2-C_{23}^2}, \quad
    \gamma_U= 0,
\end{align}
and
\begin{align}
    \alpha_\Omega = 0,\quad 
    \beta_\Omega =0, \quad
    \gamma_\Omega = 
    \frac{K^{FU}_2K^{ME}_{23}+C_{23}K^{FE}_2}{K^{FU}_2K^{M\Omega}_2-C_{23}^2}
    .
\end{align}
When the origin of $\mathbf{r}$ is set at the center of the sphere, we have $\beta_U=K^{FE}_{2}/K^{FU}_2$ and $\gamma_\Omega=K^{ME}_{23}/K^{M\Omega}_2$.
Under this symmetry, the coupling mobility tensors take the form of 
\begin{align}
    \bm{\mathcal{M}}_{12}
    =
    -
    \frac{C_{23}}{K^{FU}_2K^{M\Omega}_2-C_{23}^2} 
    (\mathbf{e}_2\mathbf{e}_3-\mathbf{e}_3\mathbf{e}_2).
\end{align}
When $C_{23}\neq0$, there exists the propulsive translation-rotation coupling for micro-swimmers, while it vanishes when the center of the resistance matches the reference point.

\subsection{Heterochiral fluid}
\label{sec:symmetry2}

We consider a fluid with a class of uniaxial symmetry in the presence of an additional reflection symmetry across a plane perpendicular to the symmetry axis [figure \ref{fig:symmeties}(c)].
The fluid thus remains invariant under $\mathbf{e}\to-\mathbf{e}$, and the corresponding point group is $C_{\infty h}$.
Such a system has previously been discussed in the classification of three-dimensional anisotropic viscosity tensors \citep{khain2022}.
Following the terminology introduced for helicoidal particles \citep{ishimoto2020jeffery}, we refer to the fluid with this class of symmetry as a heterochiral fluid.

Under the symmetry consideration as above, we have $K^{F\Omega}_{1}=K^{F\Omega}_{2}=K^{MU}_{1}=K^{MU}_{2}=C_{23}=K^{FE}_1=K^{FE}_2=K^{FE}_{23}=0$.
The strain-induced translational coefficients vanish
\begin{align}
    \alpha_U =0,\quad
    \beta_U  = 
    0, \quad
    \gamma_U= 0,
    \end{align}
whereas the corresponding angular-velocity coefficients can remain nonzero
    \begin{align}
    \alpha_\Omega = \frac{K^{ME}_1}{K^{M\Omega}_1},
    \quad
    \beta_\Omega =
    \frac{K^{M\Omega}_2K^{ME}_{2}-K^{M\Omega}_{23}K^{ME}_{23}}{(K^{M\Omega}_2)^2+(K^{M\Omega}_{23})^2},
    \quad
    \gamma_\Omega = 
    \frac{K^{M\Omega}_2K^{ME}_{23}+K^{M\Omega}_{23}K^{ME}_{2}}{(K^{M\Omega}_2)^2+(K^{M\Omega}_{23})^2}
    \label{eq:hetero}
    .
\end{align}
The coefficients $\alpha_\Omega$ and $\beta_\Omega$ are parity-odd and $\gamma_\Omega$ is parity-even, meaning that $\alpha_\Omega$ and $\beta_\Omega$ represent chirality in the fluid, while $\gamma_\Omega$ results from the fluid anisotropy.
This behavior is consistent with the exact solution for a rotating sphere in a fluid with an odd viscosity, which satisfies the identical symmetry group of a heterochiral fluid (see appendix~\ref{app} for the comparison with the exact results).
It is worth noting that even without the odd component $(K^{M\Omega}_{23}=0)$, one may find the nonzero values of $\beta_\Omega=K^{ME}_2/K^{M\Omega}_2$ and $\gamma_\Omega=K^{ME}_{23}/K^{M\Omega}_2$.
Thus, structurally chiral fluids without odd components can still exhibit a shear-induced transverse rotation along $\mathbf{q}_2,\mathbf{q}_3$.
Note that the relative translational velocity vanishes in this case $(\mathbf{U}-\mathbf{U}^\infty=\mathbf{0})$, meaning that for $\mathbf{U}^\infty\neq\mathbf{0}$, the passive particle is advected by the background flow without an additional strain-induced translational velocity while undergoing the orientational dynamics determined by the nonvanishing parameters in equation~\eqref{eq:hetero}.

Under this symmetry, the coupling mobility tensors vanish $\bm{\mathcal{M}}_{12}=\mathbf{0}$.
The vanishing off-diagonal mobility blocks eliminate force-rotation and torque-translation coupling. 
Further separation of surface-actuation modes requires an analysis of the surface traction fields.

\subsection{Homochiral fluid}
\label{sec:symmetry3}

With the $\pi$-rotational symmetry around an axis perpendicular to the axis of symmetry $\mathbf{e}$, we find $K^{FU}_{23}=K^{M\Omega}_{23}=C_{23}=K^{FE}_1=K^{FE}_2=K^{ME}_1=K^{ME}_2=0$.
We refer to the fluid with this class of symmetry as a homochiral fluid, following the terminology employed for helicoidal particles \citep{ishimoto2020jeffery}.
For the fluid satisfying this symmetry, as shown in figure \ref{fig:symmeties}(d), the coefficients are found as
\begin{align}
    \alpha_U = 0,\quad
    \beta_U  =0, \quad
    \gamma_U= 
    \frac{K^{M\Omega}_2K^{FE}_{23}
    -
    K^{MU}_2K^{ME}_{23}}{K^{FU}_2K^{M\Omega}_2
    -K^{F\Omega}_2K^{MU}_2},
\end{align}
and
\begin{align}
    \alpha_\Omega = 0,\quad
    \beta_\Omega  =0, \quad
    \gamma_\Omega= 
    \frac{K^{FU}_2K^{ME}_{23}
    -
    K^{F\Omega}_2K^{FE}_{23}}{K^{FU}_2K^{M\Omega}_2
    -
    K^{F\Omega}_2K^{MU}_2}
    .
\end{align}
The coefficients $\gamma_U$ and $\gamma_\Omega$ can take either negative or positive values, depending on the magnitude of the resistance coefficients.
Under the reflection transformation in a plane containing the symmetry axis $\mathbf{e}$, $\gamma_U$ is parity-odd, whereas $\gamma_\Omega$ is parity-even.

Under this symmetry, the coupling mobility tensors take the form of 
\begin{align}
    \bm{\mathcal{M}}_{12}
    =
     \frac{K^{F\Omega}_1}{K^{F\Omega}_1K^{MU}_1
     -K^{FU}_1K^{M\Omega}_1}
    \mathbf{ee}
    +
    \frac{K^{F\Omega}_2}{K^{F\Omega}_2K^{MU}_2
    -K^{FU}_2K^{M\Omega}_2}
    (\mathbf{I-ee})
    .
\end{align}
We observe that fluid chirality induces the translation-rotation coupling for swimming behavior, which has an anisotropic response parallel and perpendicular to the axis $\mathbf{e}$.

\begin{figure}[htbp]
    \centering
    \includegraphics[width=.8\textwidth]{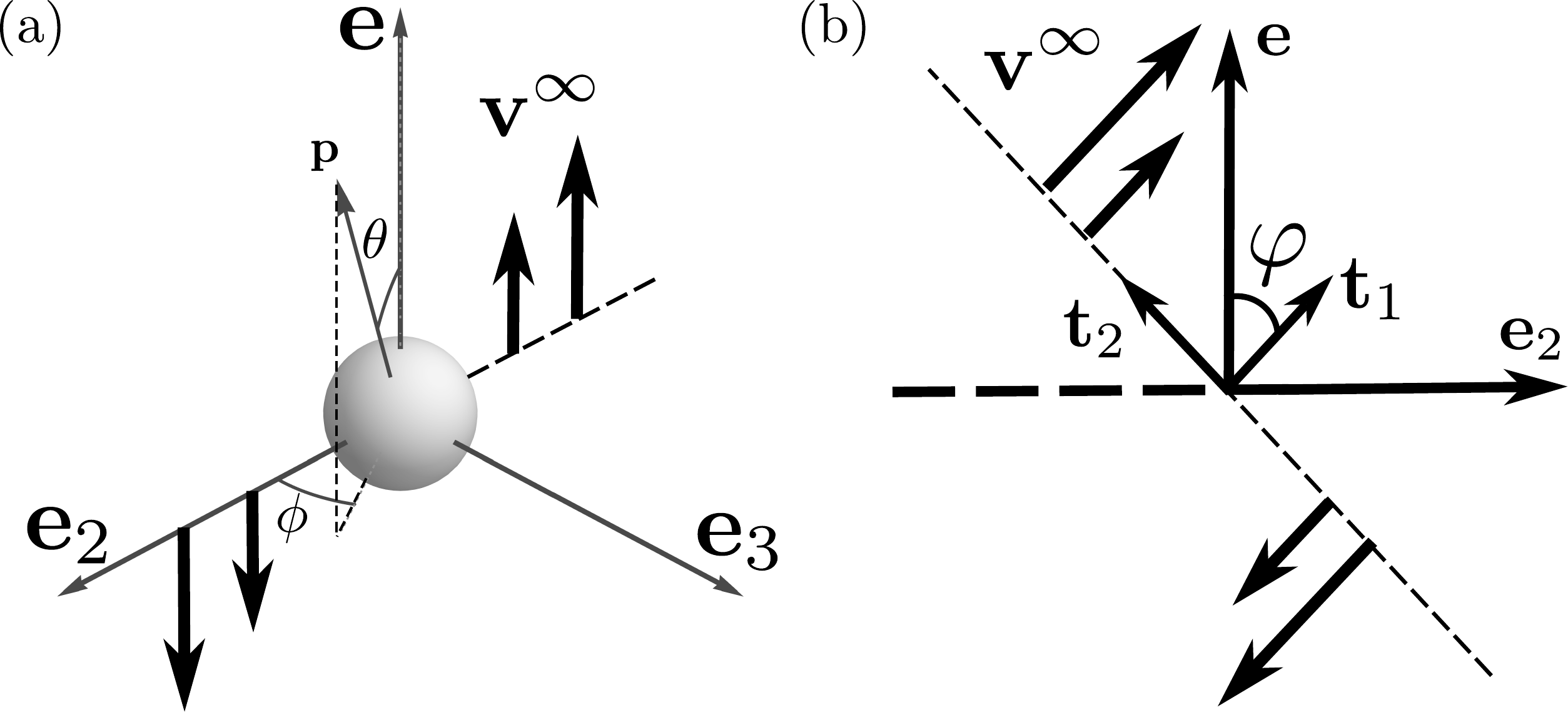}
    \caption{
Schematics of a spherical body under a background simple shear flow in the laboratory frame $\{\mathbf{e},\mathbf{e}_2,\mathbf{e}_3\}$ with $\mathbf{e}$ being the symmetry axis of a chiral uniaxial fluid.
A simple shear is considered in the $(\mathbf{e},\mathbf{e}_2)$-plane, $\mathbf{v}^\infty=\dot{\gamma}\mathbf{t}_1\mathbf{t}_2\cdot\mathbf{r}$ where $\dot{\gamma}$ is the shear strength and the flow and velocity-gradient directions are defined by unit vectors $\mathbf{t}_1=\cos\varphi\,\mathbf{e}+\sin\varphi\mathbf{e}_2$ and $\mathbf{t}_2=\sin\varphi\mathbf{e}-\cos\varphi\mathbf{e}_2$.
The corresponding background vorticity vector is along the $+\mathbf{e}_3$-axis.
(a) A spherical body under a simple shear flow when $\varphi=0$.
The orientation of the sphere with the body-fixed axis $\mathbf{p}$ is described by the polar angle $\theta$ and the azimuthal angle $\phi$ in the spherical coordinates when the axis of symmetry $\mathbf{e}$ is selected as a positive polar direction.
(b) The in-plane angle $\varphi$ specifies the direction of the shear velocity $\mathbf{t}_1$ and its spatial gradient $\mathbf{t}_2$.
    }
    \label{fig:passive}
\end{figure}

\section{Dynamics of a spherical particle in a simple shear}
\label{sec:passive}

\subsection{Passive spherical body}

To examine the shear-induced motion of a passive spherical body, we set \(\mathbf U^{prop}=\boldsymbol\Omega^{prop}=\mathbf0\). 
Since shear flow and fluid anisotropy are coupled only when $\mathbf{E}^\infty\cdot\mathbf{e}\neq\mathbf{0}$, we vary the orientation of the imposed simple shear within the $(\mathbf{e},\mathbf{e}_2)$-plane.
Its flow and velocity-gradient directions are defined by unit vectors, $\mathbf{t}_1=\cos\varphi\,\mathbf{e}+\sin\varphi\mathbf{e}_2$ and $\mathbf{t}_2=\sin\varphi\mathbf{e}-\cos\varphi\mathbf{e}_2$, respectively, where the in-plane polar angle $\varphi$ is measured from the axis $\mathbf{e}$, as schematically shown in figure~\ref{fig:passive}.
The background flow is then $\mathbf{v}^\infty=\dot\gamma \mathbf{t}_1\mathbf{t}_2\cdot\mathbf{r}$, where the constant $\dot{\gamma}$ is the shear strength.
The resulting rate-of-strain tensor and vorticity field of the background flow are 
\begin{align}
\mathbf E^\infty
=
\frac{\dot\gamma}{2}
\left[
\sin(2\varphi)
\left(\mathbf e\mathbf e-\mathbf e_2\mathbf e_2\right)
-
\cos(2\varphi)
\left(\mathbf e\mathbf e_2+\mathbf e_2\mathbf e\right)
\right],\quad
 \bm{\Omega}^\infty
    =
    \frac{\dot{\gamma}}{2}\mathbf{e}_3.
\end{align}
While the background rate-of-strain tensor has components parallel and perpendicular to the $\mathbf{e}$-direction, the vorticity vector is directed along the $\mathbf{e}_3$-axis.

Inserting the above expressions into equations~\eqref{eq:u_sym} and \eqref{eq:omega_sym} yields the following forms of the translational and rotational velocities induced by the simple shear flow:
\begin{align}
    \mathbf{U}^{shear} &=
    \frac{\dot \gamma}{2}
    \left[
    \alpha_U 
    \sin(2\varphi)
    \mathbf{e}
    -
    \cos(2\varphi)
    (
    \beta_U
    \mathbf{e}_2
    -
    \gamma_U
    \mathbf{e}_3
    )\right],\\
    \bm\Omega^{shear} &=
        \frac{\dot \gamma}{2}
        \left[
    \alpha_\Omega
    \sin(2\varphi)
    \mathbf{e}
    -
    \cos(2\varphi)
    \beta_\Omega
    \mathbf{e}_2
    +
    \{1
    +
    \cos(2\varphi)
    \gamma_\Omega
    \}
    \mathbf{e}_3\right].
    \label{eq:Oshear}
\end{align}
When $\varphi=n\pi/2$ ($n \in \{0, 1, 2, 3, \dots\}$), the velocities along $\mathbf{e}$ vanish.
These configurations include alignment of $\mathbf{e}$ either along the flow direction, $\mathbf{t}_1$ or along the velocity-gradient direction $\mathbf{t}_2$.
Conversely, for the configurations $\varphi=\pi/4+n\pi/2$, the transverse components of the velocities due to the fluid anisotropy vanish, leaving only the terms of $\alpha_U$ and $\alpha_\Omega$.

When $\varphi=0$, for instance, a sphere in an anisotropic fluid undergoes a drift velocity in a direction within the $(\mathbf{e}_2,\mathbf{e}_3)$-plane, whose angle is determined by a unit vector $(-\beta_U\mathbf{e}_2+\gamma_U\mathbf{e}_3)/\sqrt{(\beta_U)^2+(\gamma_U)^2}$.
In a transversely isotropic fluid, the sphere is drifted along the $\mathbf{e}_2$-axis as $\gamma_U=0$, and in a homochiral fluid, it is transported to $\mathbf{e}_3$, while the particle in a heterochiral fluid does not undergo any motion relative to the background shear flow as $\beta_U=\gamma_U=0.$

The rotational dynamics, on the other hand, is qualitatively similar for transversely isotropic and homochiral fluids, and the vorticity about $\mathbf{e}_3$ is modified, as $\alpha_\Omega=\beta_\Omega=0$.
In contrast, heterochiral fluids endow a particle with additional rotation about $\mathbf{e}$ and $\mathbf{e}_2$, while they do not affect the particle's positional trajectory, as $\alpha_U=\beta_U=\gamma_U=0$ in this case.

Here we discuss the typical dynamics of a particle in an external shear flow, and investigate how the fluid chirality affects the positional trajectories.
By denoting the particle center located at $\mathbf{R}(t)$, its dynamics obeys $\dot{\mathbf{R}}(t) = \dot{\gamma}\mathbf{t}_1\mathbf{t}_2\cdot\mathbf{R}(t)+\mathbf{U}^{shear}$.
The explicit expressions for the particle displacement are obtained as
\begin{align}
\mathbf{R}(t)
=
\mathbf{R}(0)+
[
\dot{\gamma}\mathbf t_1 \mathbf t_2\cdot\mathbf{R}(0)
+\mathbf U^{ shear}
]t
+
\frac{\dot\gamma^2t^2}{4}
\cos\varphi
\left[
2\alpha_U\sin^2\varphi
+
\beta_U\cos(2\varphi)
\right]\mathbf{t}_1.
\end{align}
Since the flow and velocity-gradient directions are orthogonal, the projection of the particle velocity onto the velocity-gradient direction is constant. Integrating this component gives a transverse displacement linear in time.
Substituting it back into the equation of motion reveals a velocity that varies linearly with time, whose subsequent integration produces the quadratic contribution to the positional trajectory, as in the third term above.
Consequently, the strain-translation coefficients, $\alpha_U$ and $\beta_U$, induce motion along the shear direction, which has a distinctive time-dependence.

\subsection{Active spherical body in a heterochiral fluid}

Here we consider the behavior of a spherical swimmer with nonvanishing propulsive velocities, $\mathbf{U}^{prop}\neq\mathbf{0}$ and $\bm{\Omega}^{prop}\neq\mathbf{0}$.
To obtain these swimming velocities, one needs to evaluate the surface integral with the active surface velocity profile and the auxiliary traction field.
Therefore, symmetry arguments alone do not determine the propulsion velocities of active swimmers.

In the context of chiral fluids, fluids with a single even (shear) viscosity and a single odd viscosity have been studied, as such models provide some of the simplest descriptions of chiral fluids.
These media have the continuous rotational symmetry about a preferred direction with reflection symmetry across the plane perpendicular to it \citep{khain2022}, which has the same symmetries as a heterochiral fluid discussed in section~\ref{sec:symmetry2}.
Under the incompressibility condition, imposing these constraints and assuming that the resulting three shear viscosities are all equal and the two odd viscosities are proportional to each other as assumed in previous studies \citep{khain2022, yuan2023stokesian, everts2024dissipative, lier2024odd}, leads to the viscosity tensor $\bm{\eta} = \bm{\eta}^{\rm e}+ \bm{\eta}^{\rm o}$, whose even and odd components are characterized by a single shear viscosity coefficient $\eta^{\rm e}$ and a single odd viscosity coefficient $\eta^{\rm o}$:
\begin{align}
    \eta^{\rm e}_{ijk\ell}
    &=
    \eta^{\rm e}
    \left(
    \delta_{ik}\delta_{j\ell}+
    \delta_{i\ell}\delta_{jk}
    -
    \frac{2}{3} \delta_{ij}\delta_{k\ell}
    \right),
   \\
    \eta^{\rm o}_{ijk\ell}
    &=
    \frac{    \eta^{\rm o}}{2}
    {e}_q
    (
    \epsilon_{qik}
    \delta_{\ell j}
    +
    \epsilon_{qi\ell}
    \delta_{kj}
    +
    \epsilon_{qjk}
    \delta_{\ell i}
    +
    \epsilon_{qj\ell}
    \delta_{ki}
    )
    .
\end{align}
Through the force-balance equation, the bulk fluid is described by the following three-dimensional Stokes equations
\begin{align}
    \eta^{\rm e}\nabla^2 \mathbf{v}-\frac{\eta^{\rm o}}{2}
    \mathbf{e}\cdot\nabla\boldsymbol{\omega}
    -\nabla P =\mathbf{0}
    ,
    \quad
    \nabla\cdot\mathbf{v}=0,
    \label{eq:stokes}
\end{align}
with the vorticity $\boldsymbol{\omega}=\nabla\times\mathbf{v}$ and the effective pressure $P=p-\eta^{\rm o}\mathbf{e}\cdot\boldsymbol{\omega}$.

In an odd-viscous fluid governed by equations~\eqref{eq:stokes}, the swimming velocities of a spherical squirmer were derived analytically in previous studies \citep{hosaka2024chirotactic, chatzittofi2026self}.
Using the exact relations for the traction field of an auxiliary problem, one can find from equation~\eqref{eq:UMprop} the swimming speed for any slip velocity profile.
While the translational speed takes the identical expression as that in a Newtonian fluid \citep{stone1996propulsion}, the rotational one acquires additional contributions for finite dipolar-surface modes \citep{hosaka2024chirotactic}.
For an axisymmetric squirmer, the nonvanishing swimming speeds are therefore possible when the source-dipole (characterized by a $B_1$ squirming mode) and force-dipole (characterized by a $B_2$ squirming mode) surface modes are present, which, respectively, lead to the translational and rotational motions.
For a swimmer with the radius $a$ and director vector $\mathbf{p}$, whose effective surface slip velocity is given by 
\begin{align}
\mathbf{v}^s = -B_1 (\mathbf{I}-\mathbf{nn})\cdot\mathbf{p}
+B_2  (\mathbf{p}\cdot\mathbf{n})(\mathbf{I}-\mathbf{nn})\cdot\mathbf{p},
\end{align}
the swimming speeds are obtained as \citep{hosaka2024chirotactic, chatzittofi2026self}
\begin{align}
    \mathbf{U}^{prop}&=\frac{2}{3}B_1\mathbf{p},\\
    \bm{\Omega}^{prop}&=
    \frac{3g_1B_2}{5a}(\mathbf{e\cdot p})\mathbf{p}
    -\frac{\lambda B_2}{20a}\left[1-3g_2 (\mathbf{e}\cdot\mathbf{p})^2\right]\mathbf{e}
    +
   \frac{3g_2B_2}{5a}(\mathbf{e\cdot p})\mathbf{e\times p}
    .
\end{align}
In the above, we have defined the scaling function of the odd-to-even viscosity ratio $\lambda=\eta^{\rm o}/\eta^{\rm e}$ as
\begin{align}
    g_n=
    \frac{(\lambda/4)^n}{1+(\lambda/4)^2} \quad(n=0,1,2).
\end{align} 
The sign of the $B_2$ squirming mode distinguishes between the types of microswimmers with two different propulsion mechanisms: $B_2 < 0$  corresponds to pushers, which are pushed from behind, while $B_2>0$ corresponds to pullers, which are pulled forward \citep{lauga2020fluid}.

We examine in this section how the background flow affects the orientational dynamics of a microswimmer. 
The net evolution of the swimming direction in the presence of the external shear can be written as $\dot{\mathbf{p}}=(\bm\Omega^{shear}+\bm\Omega^{prop})\times\mathbf{p}$.
For the background flow, we assume a vertical simple shear flow, $\mathbf{v}^\infty=-\dot{\gamma}(\mathbf{e}_2\cdot\mathbf{r})\mathbf{e}$, whose vorticity axis is perpendicular to the chiral axis $\mathbf{e}$.
The director dynamics can be described by the spherical coordinates with the polar and azimuthal angles $(\theta,\phi)$, and the director vector $\mathbf{p}=p_1\mathbf{e}+p_2\mathbf{e}_2+p_3\mathbf{e}_3$ is parameterized as
$p_1=  \cos\theta, p_2=\sin\theta\cos\phi$, and $p_3=\sin\theta\sin\phi$.
Noting that $\dot{p}_1=-\dot\theta \sin\theta$ and $p_2\dot p_3-p_3\dot p_2=\dot\phi \sin^2\theta$, we may rewrite the time evolution of the director vector as
\begin{align}
    \dot\theta &= 
    \frac{1}{2}\dot{\gamma} 
    ( 
    g_0 \cos\phi
    +
    g_1\sin\phi
    )
    +\frac{3g_2B_2}{5a}\sin\theta\cos\theta,
    \label{eq:thetadot}
\\
    \dot\phi &= 
    -\frac{\lambda B_2}{20a}(1-3g_2 \cos^2\theta)
    +
    \frac{1}{2}\dot{\gamma}
    \frac{g_1\cos\phi-g_0\sin\phi}{\tan\theta}
    .
    \label{eq:phidot}
\end{align}
See appendix~\ref{app:angular} for the general case where the direction of the shear flow is not aligned with the axis of symmetry $\mathbf{e}$, i.e., $\varphi\neq0$.

\begin{figure}[htbp]
    \centering
    \includegraphics[width=0.95\textwidth]{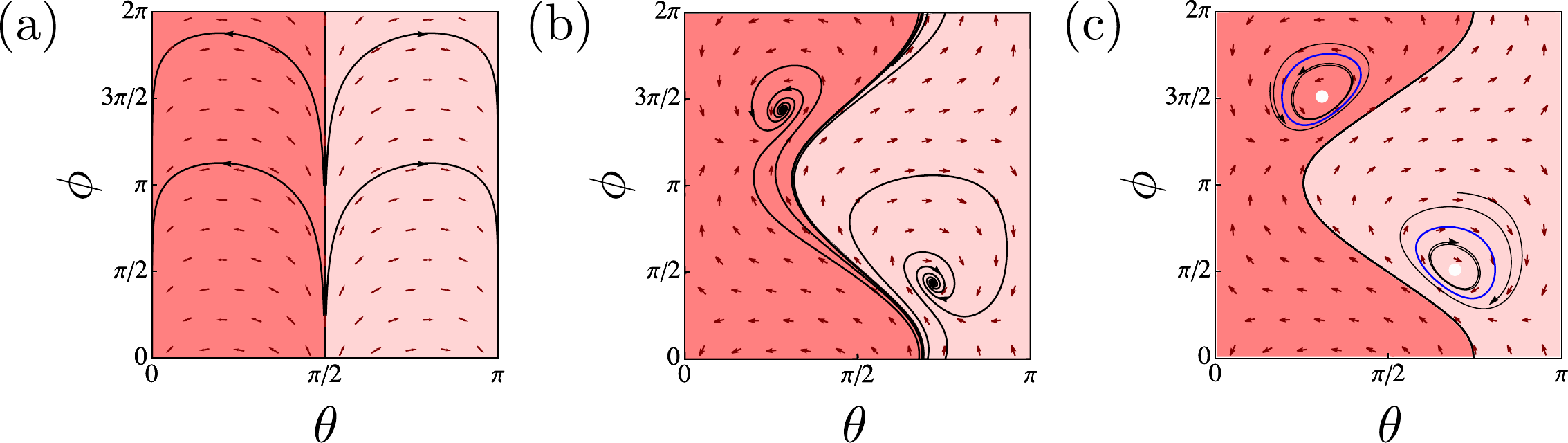}
    \caption{
Phase portraits of the orientational dynamics of a pusher $(B_2/B_1=-1)$ for different values of the shear rates (a) $\dot{\gamma}=0$, (b) $\dot{\gamma}(a/B_1)=0.5$, and (c) $\dot{\gamma}(a/B_1)=0.58$, while keeping $\lambda=4$.
When $\dot{\gamma}=0$ [panel (a)], the phase space has stable trajectories along the lines at $\theta=0,\pi$, characteristic of bimodal chirotaxis.
An unstable limit cycle is along the line at $\theta=\pi/2$, acting as a separatrix.
Below the critical shear rate $\dot{\gamma}<\dot{\gamma}_c$ [panel (b)], the phase space contains two spiral sinks (black circles), corresponding to bimodal chirotactic alignment.
With increasing $\dot{\gamma}$, the stable fixed points lose stability through a supercritical Hopf bifurcation at $\dot{\gamma}=\dot{\gamma}_c$.
For $\dot{\gamma}>\dot{\gamma}_c$ [panel (c)], the two fixed points become unstable spirals (white circles), and two stable limit cycles emerge (blue).
The basins of attraction of the two stable states are shaded in red and pale red, and are separated by a separatrix, an unstable limit cycle.
The azimuthal coordinate is shifted according to $\phi\to\phi+\pi$ to place the fixed points away from the boundaries of the plotting window.
}
    \label{fig:Fig5}
\end{figure}

We first recapitulate the typical dynamics of a microswimmer without the background flow, whose dynamical system in the $(\theta,\phi)$ plane is plotted in figure \ref{fig:Fig5}(a), with the odd-to-even viscosity ratio fixed to be $\lambda=4$.
In this case, the microswimmer exhibits the chirotactic response, characteristic in three-dimensional odd fluids \citep{hosaka2024chirotactic}.
When $\dot{\gamma}=0$,  the director dynamics in equations~\eqref{eq:thetadot} and \eqref{eq:phidot} is determined solely by the polar angle $\theta$ and is independent of the azimuthal angle $\phi$.
This is because the system has a continuous rotational symmetry about the axis $\mathbf{e}$.
For the pusher-type swimmer, the stable angles are given by $\theta=0,\pi$, meaning that the swimmer undergoes precession around the axis $\mathbf{e}$, and eventually exhibits bistable behavior reorienting itself either to $+\mathbf{e}$ or $-\mathbf{e}$.
Such a chirality-induced reorientation is referred to as chirotaxis, and the bistable case for the axisymmetric swimmer is specifically called bimodal chirotactic response \citep{hosaka2024chirotactic}.
Note that the leading alignment effects are of $\mathcal{O}(\lambda^2)$ [the second term on the right-hand side of equation~\eqref{eq:thetadot}], while the precession is $\mathcal{O}(\lambda)$ [the first term on the right-hand side of equation~\eqref{eq:phidot}].

To investigate the effect of the background shear on the stable orientational behavior, we study the phase portraits when $\dot{\gamma}\neq0$ [figures \ref{fig:Fig5}(b,c)], while keeping $\lambda=4$, as before.
At low shear rates [figure~\ref{fig:Fig5}(b)], the system has two spiral sinks that are characteristic of bimodal chirotaxis.
When $\dot{\gamma}$ is increased, the two sinks lose stability, and above the bifurcation, two stable limit cycles emerge around the unstable fixed points [figure~\ref{fig:Fig5}(c)].
This indicates the breakdown of the stable orientational behaviors of bimodal chirotaxis.
In the parameter regime considered, the transition is classified as a supercritical Hopf bifurcation, and the corresponding critical shear rate can be derived as $\dot{\gamma}_c=|\lambda B_2|/(5\sqrt{2}a)$.
As $\lambda$ is increased, the critical value also increases, implying that larger shear rates are needed to overcome the aligning effect due to the fluid chirality.

These orientational regimes have distinct consequences for swimmer trajectories. 
Below the critical shear rate [figure \ref{fig:Fig5}(a,b)], the director approaches one of the two stable orientations.
When stable alignment is lost [figure \ref{fig:Fig5}(c)], the swimming direction can undergo persistent oscillatory rotation about the unstable fixed points.
Importantly, the external shear preserves the bimodal alignment of the axisymmetric pusher-type swimmer under sufficiently small shear rates.
Moreover, shear-induced destabilization of bimodal chirotaxis does not, by itself, generate unimodal chirotaxis, where a chiral swimmer reorients itself along a single direction \citep{hosaka2024chirotactic}.
Instead, the main effect of sufficiently strong shear is the loss of stable alignment, and the emergence of time-dependent oscillatory  orientational dynamics.

\section{Discussion and conclusions}
\label{sec:discussion}

We have developed a general theoretical framework for the hydrodynamics of microswimmers in fluids governed by linearly viscous constitutive relations, encompassing orientational anisotropy, chirality, and parity-violating odd viscosities. 
By employing the generalized Lorentz reciprocal theorem, we established the Onsager-Casimir reciprocity for the resistance tensors of an arbitrarily shaped rigid body. 
We further derived a swimming formula that accounts for both prescribed surface actuation and imposed background flows.
Using this framework, we investigated the motion of passive spherical particles and active spherical swimmers under an external shear flow in chiral uniaxial fluids.
Our symmetry analysis identified the distinct forms of shear-induced translational and rotational responses compatible with transversely isotropic, heterochiral, and homochiral symmetry classes.
We also examined how an imposed shear flow affects the bistable orientational alignment of spherical swimmers, known as bimodal chirotaxis, in a fluid with odd viscosity.
Our stability analysis revealed that sufficiently strong shear can destabilize the steady swimming orientations, leading to the breakdown of stable chirotactic alignment.
These results serve as a systematic theoretical basis for understanding the dynamics of passive particles and active microswimmers in chiral and anisotropic viscous environments, particularly their responses to externally imposed flows.

The derivation of a generalized swimming formula enabled us to prove the fundamental principle for low-Reynolds-number swimming, and show that the scallop theorem remains valid within the framework of the linear constitutive relation.
The underlying fluid properties govern the tensorial structure of the mobility, but as long as they remain spatiotemporally constant and independent of the instantaneous flow rate, the kinematics remains perfectly time-reversible. 
Thus, the scallop theorem holds robustly in any generalized linearly viscous fluid.
The consequence of the reciprocal motion in such media is consistent with the analysis of an oscillating dumbbell-shaped dimer molecule \citep{Chatzittofi2025}.
Note that the scallop theorem was established using the geometric theory for two-dimensional incompressible fluids with an isotropy-compatible odd viscosity \citep{lapa2014}.

The derived equations governing the translational and orientational dynamics of a spherical swimmer reveal previously unexplored dynamical regimes, including anomalous shear-induced transverse migration and orientational trapping in external flows. 
These nontrivial fluid-particle interactions provide a theoretical foundation for investigating swimmers with more complex geometries and may offer insights into microbial transport through structured biological fluids, as well as providing the designing principles of artificial active matter systems that exploit chiral and anisotropic fluid rheology.
Previous studies of magnetically driven particles with different shapes have demonstrated that particle anisotropy can influence propulsion velocities \citep{morozov2017dynamics,chen2021propulsion}. 
It would therefore be interesting to investigate how particle shape anisotropy couples with the chirality and anisotropy of the surrounding fluid, and how this coupling modifies the resulting translational and orientational dynamics. 
We leave this question for future investigation.

Although the motion of spherical particles in fluids with a single odd viscosity coefficient has been extensively studied, its generalization to fluids with two independent odd viscosity coefficients remains relatively unexplored, despite recent progress in this direction \citep{lier2026sorting}. 
Our symmetry-based framework therefore provides a useful theoretical approach to investigating particle dynamics in such fluids with two independent odd-viscosity coefficients.
While symmetry constraints on translational and rotational responses have previously been examined \citep{khain2024trading}, the response to an imposed background flow additionally depends on shear-induced resistance tensors, whose symmetry-allowed structures have been systematically analyzed in this work. 
Beyond the spherical particles considered here, a natural direction for future research is to extend this framework to anisotropic and chiral particles, for which the interplay between particle geometry and fluid anisotropy may give rise to richer dynamical behavior, particularly in fluids with two independent odd viscosity coefficients and in fluids with more general symmetry classes.

\begin{acknowledgements}
We thank B.\ Mahault and A.\ Vilfan for fruitful discussions.
Y.H.\ and K.I.\ acknowledge support from the Japan Science and Technology Agency (JST) CREST (Grant No.\ JPMJCR25Q1).
K.I.\ acknowledges the Japan Society for the Promotion of Science (JSPS) KAKENHI (Grant No.\ 24K21517) and the Japan Science and Technology Agency (JST), FOREST (Grant No.\ JPMJFR212N). 
This work was supported by the Japan Society for the Promotion of Science (JSPS) Core-to-Core Program ``Advanced core-to-core network for the physics of self-organizing active matter'' (JPJSCCA20230002).
\end{acknowledgements}

\appendix

\section{Angular dynamics of a spherical body in a fluid with odd viscosity}
\label{app}

Here we derive an exact analytical expression of the shear-induced angular dynamics of a spherical body in a three-dimensional fluid with a single odd viscosity coefficient, as described in equation~\eqref{eq:stokes}.
Although the symmetry arguments we considered do not impose a restriction on the number of odd viscosities in chiral uniaxial fluids, we focus here on the special case where the problem is tractable, namely, a fluid with a single odd viscosity.
This enables us to compare the results based on the symmetry arguments and those by analytical calculations.

For a spherical body moving in the odd Stokes equation~\eqref{eq:stokes}, there are no coupling terms between the force and torque as well as the shear-translation coupling, i.e., $\mathbf{\hat{K}}^{F\Omega}=\mathbf{\hat{K}}^{MU}=\mathbf{0}$ and $\mathbf{\hat{K}}^{FE}=\mathbf{0}$ \citep{everts2024dissipative}.
The time evolution of the orientational dynamics of a sphere follows from equation~\eqref{eq:UOmega2}
\begin{align}
    \bm{\Omega}
    =
    \bm{\Omega}^\infty
    +
    [(\mathbf{\hat{K}}^{M\Omega})^\top]^{-1}\cdot\mathbf{\hat{K}}^{ME}:
    \mathbf{E}^\infty.
    \label{eq:Omegai}
\end{align}
To obtain the director equation, we need to obtain the second-rank rotational-torque tensor, $\mathbf{\hat{K}}^{M\Omega}$, and the third-rank shear-induced resistance tensor, $\mathbf{\hat{K}}^{ME}$, which can be derived from a solution of a rotating sphere in an odd-viscous fluid.

One strategy for solving the odd Stokes equations analytically is to map them into the classical ones without odd viscosity, when the solutions can be described solely by pressure-free flows \citep{hosaka2024chirotactic}.
Under such a constraint, the solutions can be expressed by classical hydrodynamic singularities, solutions that do not depend on odd viscosity.
For a rotating spherical body of radius $a$ with angular velocity $\bm{\hat{\Omega}}$, the solution is therefore expressed as a rotlet singularity, $\mathbf{\hat{v}}=(a^3/r^3) \boldsymbol{\hat{\Omega}}\times\mathbf{r}$, as in a classical Newtonian fluid \citep{happel2012low}, even in the presence of odd viscosity.
The corresponding strain-rate tensor is $\mathbf{\hat{E}}=3 a^3/(2r^5)[(\mathbf{r}\times\boldsymbol{\hat{\Omega}})\mathbf{r}+\mathbf{r}(\mathbf{r}\times\boldsymbol{\hat{\Omega}})]$.
Noting that the velocity field is independent of odd viscosity, but the stress tensor is not, the traction of a rotating sphere is found as \citep{hosaka2024chirotactic}
\begin{align}
    \hat{f}_i
    &=
    3
    \bigg(
    \hat{\eta}^{\rm e}
    \epsilon_{ijk}n_k
    +
    \frac{1}{2}
    \hat{\eta}^{\rm o}    
    \epsilon_{i\ell m}
    (\epsilon_{\ell j k}
    n_k)
    e_m
    +
    \frac{1}{3}
    \hat{\eta}^{\rm o}  
    n_ie_j
    \bigg)
    \hat{\Omega}_j
    .
\end{align}
From the relation $\hat{f}_i=-\hat{\Sigma}_{ij}^\Omega\hat{\Omega}_j$, the rotational surface force resistance tensor is then given by
\begin{align}
    \hat{\Sigma}_{ij}^\Omega
    =
    -
    3
    \bigg(
     \hat{\eta}^{\rm e}
    \epsilon_{ijk}n_k
    +
    \frac{1}{2}
     \hat{\eta}^{\rm o}
    \epsilon_{i\ell m}
    (\epsilon_{\ell j k}
    n_k)
    e_m
    +
    \frac{1}{3}
     \hat{\eta}^{\rm o}
    n_ie_j
    \bigg).
\end{align}

Having obtained the traction profile, we proceed to calculate the rotation- and shear-induced resistance tensors.
By integrating the first order moment of the traction over the surface, the rotational-torque resistance tensor is obtained as
\begin{align}
    \hat{K}^{M\Omega}_{ij}=
        \int_\mathcal{S}dS\,
    \epsilon_{ik\ell}r_k
    \hat{\Sigma}^\Omega_{\ell j}
    =
    8\pi a^3
    \left(
    \hat{\eta}^{\rm e}
    \delta_{ij}+
    \frac{1}{4}
    \hat{\eta}^{\rm o}
   \epsilon_{ijk}e_k
\right)
.
\end{align}
Since the velocity field is independent of odd viscosity and the stress tensor depends linearly on $\hat{\eta}^{\rm o}$, the resistance tensor depends linearly on $\hat{\eta}^{\rm o}$.
The torque-shear tensor is obtained as
\begin{align}
    \hat{K}^{ME}_{ijk}
    =
	\frac{1}{2} \int_{S} dS\, \left( r_k \hat{\Sigma}_{j i}^{\Omega}+r_j \hat{\Sigma}_{k i}^{\Omega} \right)
    &=
\pi a^3\hat{\eta}^{\rm o}
\left(
\frac{10}{3}e_i\delta_{jk}
-\delta_{ij}e_k-\delta_{ik}e_j
\right)
.
\end{align}

From condition \eqref{eq:cond}, we find the relation of the even and odd viscosities between the main and auxiliary problems as
\begin{align}
\hat{\eta}^{\rm e}=\eta^{\rm e},\quad
\hat{\eta}^{\rm o}=-\eta^{\rm o}.
\end{align}
Using these relations, the inverse of the resistance tensor and torque-shear tensor can be expressed with the viscosity coefficients in the main problem
\begin{align}
    (\mathbf{\hat{K}}^{M\Omega})^{-1}
    =
    (8\pi\eta^{\rm e}a^3)^{-1}
    \left[
   \mathbf{I}
    +
    \frac{(\lambda/4)}{1+(\lambda/4)^2}
    \bm{\epsilon}\cdot\mathbf{e}
    +
    \frac{(\lambda/4)^2}{1+(\lambda/4)^2}
(    \mathbf{ee-I})\right]
    ,
    \label{eq:inverse}
\end{align}
and
\begin{align}
    \hat{K}^{ME}_{ijk}
	=
    -
    \pi a^3 
    \eta^{\rm o}
    \left(
\frac{10}{3}e_i\delta_{jk}
-\delta_{ij}e_k-\delta_{ik}e_j
\right)
,
    \label{eq:KMEhat}
\end{align}
with the odd-to-even viscosity ratio $\lambda=\eta^{\rm o}/\eta^{\rm e}$.
Inserting equations~\eqref{eq:inverse} and \eqref{eq:KMEhat} into equation~\eqref{eq:Omegai} gives the angular velocity
\begin{align}
    \bm{\Omega}
    =
    \bm{\Omega}^\infty
    +
    \left[
    \frac{\lambda}{4}
    \mathbf{ee}
    +
     \frac{(\lambda/4)}{1+(\lambda/4)^2}(\mathbf{I}-\mathbf{ee}) 
    -
     \frac{(\lambda/4)^2}{1+(\lambda/4)^2}
    (\mathbf{e}_2\mathbf{e}_3-\mathbf{e}_3\mathbf{e}_2)
    \right]
    \cdot
    \mathbf{E}^\infty\cdot\mathbf{e}.
    \label{eq:Omega_exa}
\end{align}
Under the parity transformation $(\lambda\to-\lambda)$, the first and second terms in equation~\eqref{eq:Omega_exa} change signs, while the third term remains unchanged.
This trend coincides with the shear-induced angular velocity obtained from the symmetry argument, given in equation~\eqref{eq:omega_sym}.
The coefficients in the angular dynamics are specifically given by
\begin{align}
	\alpha_\Omega = \frac{\lambda}4,\quad
	\beta_\Omega =     \frac{(\lambda/4)}{1+(\lambda/4)^2},\quad
	\gamma_\Omega =    -\frac{(\lambda/4)^2}{1+(\lambda/4)^2}
	.
\end{align}
To linear order in $\lambda$, one finds $\beta_\Omega\approx\lambda/4$ and $\gamma_\Omega\approx0$, and the shear-induced angular velocity is proportional to $\mathbf{E}^\infty\cdot\mathbf{e}$.
The resistance tensors derived here can also be obtained using the singularity method \citep{everts2024dissipative}.

\begin{figure}[htbp]
    \centering
    \includegraphics[width=0.6\textwidth]{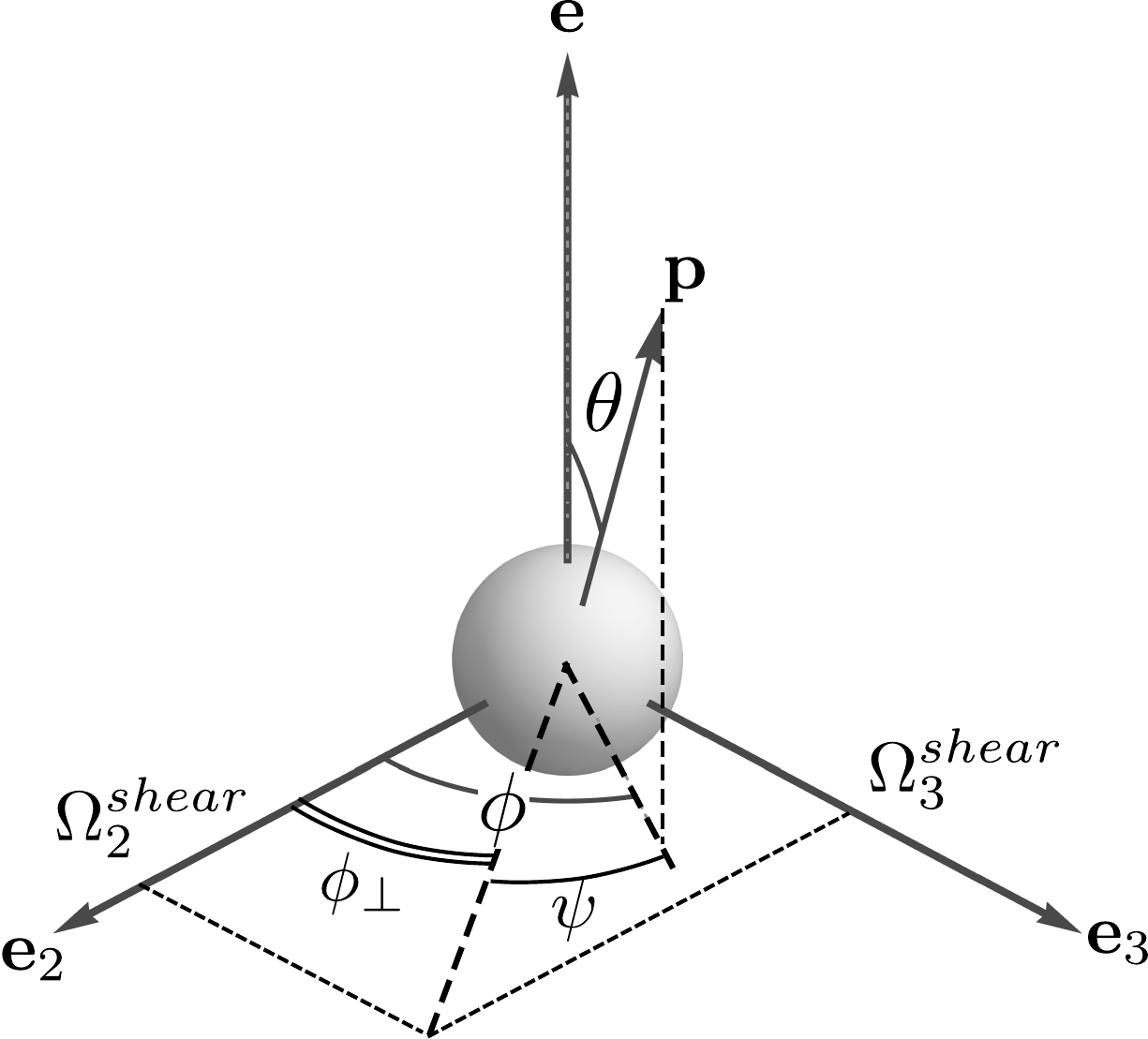}
    \caption{
Orientation of a spherical body with the body-fixed axis $\mathbf{p}$ whose polar and azimuthal angles are $\theta$ and $\phi$ in the spherical coordinates when the $\mathbf{e}$-direction is chosen as a positive polar axis.
The direction of the in-plane angular velocity $\Omega_2^{shear}\mathbf{e}_2+\Omega_3^{shear}\mathbf{e}_3$ can be specified by an angle, $\phi_\perp$, from the $\mathbf{e}_2$-axis, which redefines the azimuthal angle as $\psi=\phi-\phi_\perp$.
    }
    \label{fig:psi}
\end{figure}

\section{Angular dynamics of a spherical microswimmer in a fluid with odd viscosity}
\label{app:angular}

Here we derive the director dynamics of a spherical microswimmer when the direction of a shear flow is arbitrary in the $(\mathbf{e},\mathbf{e}_2)$ plane, i.e., the case of $\varphi\neq0$.
If we describe the director dynamics using the spherical coordinates with the polar and azimuthal angles $(\theta,\phi)$, the director vector $\mathbf{p}=p_1\mathbf{e}+p_2\mathbf{e}_2+p_3\mathbf{e}_3$ is parameterized as
$p_1=  \cos\theta, p_2=\sin\theta\cos\phi$, and $p_3= \sin\theta\sin\phi$.
Through the relations $\dot{p}_1=-\dot\theta \sin\theta$ and $p_2\dot p_3-p_3\dot p_2=\dot\phi \sin^2\theta$, we may rewrite the time evolution of the director as
\begin{align}
    \dot\theta &= \Omega_3^{shear}\cos\phi 
    -\Omega_2^{shear}\sin\phi +\frac{3g_2B_2}{5a}\sin\theta\cos\theta,
    \label{eq:appthetadot}\\
    \dot\phi &= 
        -\frac{\lambda B_2}{20a}(1-3g_2 \cos^2\theta)
     +\Omega_1^{shear}
    - (\Omega_2^{shear}\cos\phi+\Omega_3^{shear}\sin\phi)/\tan\theta,
    \label{eq:appphidot}
\end{align}
with the decomposition $\bm{\Omega}^{shear} = \Omega_1^{shear}\mathbf{e}+\Omega_2^{shear}\mathbf{e}_2+\Omega_3^{shear}\mathbf{e}_3$.
To investigate the competition between chirotactic alignment and shear-induced rotation, we decompose the external angular velocity into components parallel and perpendicular to the anisotropy axis. 
For the shear geometry as in equation~\eqref{eq:Oshear}, we find
\begin{align}
\Omega_\perp^{shear}=\frac{|\dot\gamma|}{2}
\sqrt{(\beta_\Omega)^2\cos^2(2\varphi)
+[1+\gamma_\Omega\cos(2\varphi)]^2},
\label{eq:perp}
\end{align}
for the perpendicular component, and
\begin{align}
\Omega_\|(\theta)=
    -\frac{\lambda B_2}{20a}(1-3g_2 \cos^2\theta)
+\frac{\dot\gamma}{2}\alpha_\Omega\sin(2\varphi),
\end{align}
for the parallel component.
In the above, $\Omega_\perp^{shear}$ denotes the magnitude of shear-induced rotation perpendicular to $\mathbf{e}$, while $\Omega_\|$ represents the total precessional angular velocity about the chiral axis.
The direction of the in-plane angular velocity $\Omega^{shear}_\perp$ can be specified by an angle from the $\mathbf{e}_2$-axis. 
By defining the new azimuthal angle $\psi=\phi-\phi_\perp$ as in figure~\ref{fig:psi}, the director dynamics in equations~\eqref{eq:appthetadot} and \eqref{eq:appphidot}  can be simply rewritten as
\begin{align}
\dot\theta&=-\Omega_\perp^{shear}\sin\psi+
\frac{3g_2B_2}{5a}\sin\theta\cos\theta,\\
\dot\psi&=\Omega_\|-\Omega_\perp^{shear}\cos\psi/\tan\theta,
\end{align}
where we have used the identities, $\sin\phi_\perp\cos\phi-\cos\phi_\perp\sin\phi=-\sin(\phi-\phi_\perp)$ and $\cos\phi_\perp\cos\phi+\sin\phi_\perp\sin\phi=\cos(\phi-\phi_\perp)$ and the relation $\dot{\psi}=\dot{\phi}$.

\end{document}